\documentclass[final,authoryear,3p,times]{elsarticle}
\usepackage{graphicx,color}
\usepackage{enumerate}
\usepackage[colorlinks]{hyperref}
\usepackage{amsmath,amssymb,amsthm,bm}
\usepackage{multicol}
\usepackage[dvipsnames]{xcolor}
\usepackage{ulem}
\usepackage{tikz}
\usetikzlibrary{automata,positioning}
\usepackage{afterpage}
\usepackage{calc}
\usepackage{ifthen}
\usepackage{algorithm}
\usepackage{algpseudocode}
\usepackage{caption}
\usepackage{subcaption}
\usepackage{booktabs}

\edef\svtheparindent{\the\parindent}
\usepackage{parskip}
\parindent=\svtheparindent\relax

\newcommand{\revised}[1]{{\color{black} #1}}

\newcommand{\LZ}[1]{\textcolor{black}{#1}}

\journal{CMAME}

\newcommand{\boldface}[1]{\boldsymbol{#1}}  

\newcommand{\bfb}{\boldface{b}}

\newcommand{\bfd}{\boldface{d}}
\newcommand{\bfe}{\boldface{e}}
\newcommand{\bff}{\boldface{f}}

\newcommand{\bfk}{\boldface{k}}

\newcommand{\bfn}{\boldface{n}}

\newcommand{\bft}{\boldface{t}}
\newcommand{\bfu}{\boldface{u}}

\newcommand{\bfx}{\boldface{x}}

\newcommand{\bfz}{\boldface{z}}
\newcommand{\bfA}{\boldface{A}}
\newcommand{\bfB}{\boldface{B}}

\newcommand{\bfF}{\boldface{F}}

\newcommand{\bfK}{\boldface{K}}

\newcommand{\bfS}{\boldface{S}}
\newcommand{\bfT}{\boldface{T}}
\newcommand{\bfU}{\boldface{U}}

\newcommand{\bfeps}{\boldsymbol{\varepsilon}}

\newcommand{\bfepsilon}{\boldsymbol{\varepsilon}}

\newcommand{\bfsigma}{\boldsymbol{\sigma}}

\newcommand{\bfchi}{\boldsymbol{\chi}}

\newcommand{\bfTheta}{\boldsymbol{\Theta}}

\newcommand{\calF}{\mathcal{F}}

\newcommand{\calL}{\mathcal{L}}

\newcommand{\calU}{\mathcal{U}}
\newcommand{\calV}{\mathcal{V}}

\newcommand{\calX}{\mathcal{X}}

\newcommand{\dsC}{\mathbb{C}}

\newcommand{\partderiv}[2]{\frac{\partial #1}{\partial #2}}

\newcommand{\T}{^{\mathrm{T}}} 

\newcommand{\Rset}{\mathbb{R}}

\newlength{\boxwidth}
\def\dd{\;\!\mathrm{d}}

\def\btheorem{\begin{theorem}}
	\def\etheorem{\end{theorem}}
\def\blemma{\begin{lemma}}
	\def\elemma{\end{lemma}}
\def\bproposition{\begin{proposition}}
	\def\eproposition{\end{proposition}}
\def\bcorollary{\begin{corollary}}
	\def\ecorollary{\end{corollary}}
\def\bdefinition{\begin{definition}}
	\def\edefinition{\end{definition}}
\def\bexample{\begin{example}}
	\def\eexample{\end{example}}
\def\bremark{\begin{remark}}
	\def\eremark{\end{remark}}

\def\fg{\boldsymbol}

\DeclareMathOperator{\divv}{div}

\DeclareMathOperator{\sym}{sym}

\newcommand{\be}{\begin{equation}}
\newcommand{\ee}{\end{equation}}
\newcommand{\beq}{\begin{eqnarray}}
\newcommand{\eeq}{\end{eqnarray}}
\newcommand{\bem}{\begin{multline}}
\newcommand{\eem}{\end{multline}}
\newcommand{\ba}{\begin{align}}
\newcommand{\ea}{\end{align}}
\newcommand{\bfzero}{{\fg0}}

\renewcommand{\figurename}{Figure }

\begin{document}

\begin{frontmatter}

\title{Data-driven topology optimization of spinodoid metamaterials with seamlessly tunable anisotropy}

\author[eth]{Li Zheng}
\author[eth,tud]{Siddhant Kumar}
\author[eth]{Dennis M. Kochmann\corref{cor1}}

\ead{dmk@ethz.ch}
\cortext[cor1]{Phone +41-44-632-3276.}

\address[eth]{Mechanics \& Materials Lab, Department of Mechanical and Process Engineering, ETH Z\"{u}rich, 8092 Z\"{u}rich, Switzerland}
\address[tud]{Department of Materials Science and Engineering, Delft University of Technology, Mekelweg 2, 2628 CD Delft, The Netherlands}

\begin{abstract}
	We present a two-scale topology optimization framework for the design of macroscopic bodies with an optimized elastic response, which is achieved by means of a spatially-variant cellular architecture on the microscale. The chosen spinodoid topology for the cellular network on the microscale (which is inspired by natural microstructures forming during spinodal decomposition) admits a seamless spatial grading as well as tunable elastic anisotropy, and it is parametrized by a small set of design parameters associated with the underlying Gaussian random field. The macroscale boundary value problem is discretized by finite elements, which in addition to the displacement field continuously interpolate the microscale design parameters. By assuming a separation of scales, the local constitutive behavior on the macroscale is identified as the homogenized elastic response of the microstructure based on the local design parameters. As a departure from classical FE$^2$-type approaches, we replace the costly microscale homogenization by a data-driven surrogate model, using deep neural networks, which accurately and efficiently maps design parameters onto the effective elasticity tensor. The model is trained on homogenized stiffness data obtained from numerical homogenization by finite elements. As an added benefit, the machine learning setup admits automatic differentiation, so that sensitivities (required for the optimization problem) can be computed exactly and without the need for numerical derivatives -- a strategy that holds promise far beyond the elastic stiffness. Therefore, this framework presents a new opportunity for multiscale topology optimization based on data-driven surrogate models.
\end{abstract}


\end{frontmatter}


\section{Introduction}
\label{sec:Introduction}

Supported by developments in advanced manufacturing, mechanical metamaterials with tunable microstructure and controllable properties have made significant strides towards realizing \textit{materials by design} \citep{gibson2010cellular,evans2010concepts,lee2012micro,zheng2014ultralight,meza2014,berger2017mechanical}. Despite all progress, key challenges have persisted, which can be exemplified by truss-, plate-, and shell-based cellular materials, which have dominated the design of metamaterials over the past decade. Most truss-based architectures exhibit poor scaling of stiffness and strength with relative density due to bending deformation of struts \citep{meza2017}. Plate-based designs exhibit improved stiffness with relative density than trusses and have been shown to reach upper bounds on the achievable effective stiffness of cellular media \citep{berger2017mechanical,tancogne-dejean2018}. However, both types of architectures suffer from stress localization at the junctions of beams or plates, which results in early failure and poor recoverability \citep{PortelaEtAl2018,MateosEtAl2019,LattureEtAl2018}. Smooth architectures like those based on triply-periodic minimal surfaces (TPMS) \citep{NguyenEtAl2017,HanEtAl2017,AlKetanAlRub2019} address the aforementioned issue, but they do not overcome a common limitations in the fabrication of architectures based on periodic unit cells: classically, all truss, plate, and TPMS designs produce periodic structures with high sensitivity of mechanical properties to symmetry-breaking imperfections and defects and limited opportunity for introducing smoothly spatially variant structures.

In light of these challenges, spinodal metamaterials have emerged recently as a new class of non-periodic architected \revised{material \citep{kumar2020inverse,portela2020extreme,hsieh2019mechanical,guell2019ultrahigh,vidyasagar2018microstructural,soyarslan20183d}}. Their design is inspired by topologies observed during spinodal decomposition \citep{cahn1961spinodal, ALLEN20018761}, which occurs, e.g., during rapid diffusion-driven phase separation in nanoporous alloys \citep{erlebacher2001evolution, hodge2007scaling}, polymer blends \citep{bruder1992spinodal}, and micro-emulsions \citep{lee2010bicontinuous}. The computational design of spinodal microstructures relies on simulating the phase separation process by phase field methods whose kinetics are modeled by Cahn-Hilliard-type evolution equations \citep{cahn1958free,Cook1970,stewart1992spinodal,torabi2009new,salvalaglio2015faceting,vidyasagar2018microstructural,vuijk2019effect} or, as a shortcut, by Gaussian random fields \citep{soyarslan20183d}. As a two-phase mixture spontaneously decomposes into two spatially-separated stable phases, the process is artificially arrested and solid- or shell-based topologies are extracted by, respectively, removing one of the two phases or by retaining the interfaces and eliminating both phases. The resulting topologies are composed of smooth, bi-continuous, and non-self-intersecting surfaces \revised{were shown to have intriguing properties. \cite{portela2020extreme} demonstrated that spinodal topologies exhibit better (and close to optimal) stiffness scaling with respect to relative density, and they were shown to exhibit a significantly improved mechanical resilience over comparable truss and plate metamaterials, as seen in the recovery after repeated nonlinear compression. Unlike beam and plate architectures, spinodal metamaterials (like TPMS) avoid stress concentrations at beam or plate junctions, thus reducing stress concentrations, which considerably contributes to the observed high mechanical resilience. Compared to TPMS, the non-periodicity of spinodal architectures renders them insensitive to symmetry-breaking defects and fabrication-induced imperfections \citep{hsieh2019mechanical}. \cite{guell2019ultrahigh} further demonstrated ultra-high energy absorption characteristics of spinodal architectures.} Spinodal metamaterials can also self-assemble across several length scales -- from centimeters to nanometers \citep{portela2020extreme}, which is a promising avenue to overcome the scalability challenge of additive manufacturing. \revised{(Note that in this study we do not take into account the manufacturability of structures as, e.g., \citet{watts2019simple}).}

We recently introduced a computational shortcut to generate spinodal-like topologies, referred to as \textit{spinodoid} topologies \citep{kumar2020inverse}. This approach replaces computationally expensive phase field simulations for topology generation and provides a simple parametrization based on anisotropic Gaussian random fields (GRFs) \citep{Adler2007,cahn1965phase}; unlike the isotropic formulation of \citet{soyarslan20183d}, spinodoids allow for an efficient exploration of a wide design space of anisotropic mechanical properties. When designing functionally-graded metamaterials, the GRF-based approach admits seamless, spatially-variant topologies, which in contrast to periodic unit-cell-based designs bypasses discontinuities and tessellation-related limitations.

Spinodoid metamaterials bear potential for applications ranging from energy absorption and impact protection to heat exchange and to synthetic bone. An ongoing challenge in the design of patient-specific bone implants is to match the anisotropic topological and mechanical properties of bone -- which can be highly heterogeneous across patients as well as within the same bone. Functionally-graded spinodoid metamaterials were shown to be promising candidates for inverse-designed synthetic bones \citep{kumar2020inverse} for improved biomechanical compatibility and reduced bone atrophy. Yet, aside from optimizing the properties of specific spinodoid topologies, they have not been used in any two-scale design challenge, such as, e.g., identifying the optimal macroscale shape of a bone implant while optimizing the local, spatially-varying spinodoid microstructure. To this end, we here address the systematic design of spatially-variant, functionally-graded bodies with a spinodoid microstructure through a data-driven topology optimization approach.

Topology optimization is a well-established technique (see \citet{bendsoe2013topology} and \citet{sigmund2013review} for detailed reviews). The classical Solid Isotropic Material Penalization (SIMP) method and its extensions \citep{bendsoe1988generating,bendsoe1989optimal,sigmund1998numerical,bendsoe1999material,sigmund200199} define a continuous volume fraction field $\rho:\Omega\to[0,1]$ over a body $\Omega\subset\Rset^d$ in $d$ dimensions, such that the local linear elastic modulus tensor is approximated as
\begin{equation}\label{eq:cx}
	\mathbb{C}= \mathbb{C}_0 + (\mathbb{C}_S - \mathbb{C}_0)\rho^p,
\end{equation}
where  $\mathbb{C}_0$ and $\mathbb{C}_S$ are the stiffness tensors of void and solid regions, respectively, and $p\geq 3$ is a penalization exponent to promote (approximately) purely void or solid states. Most SIMP-based methods optimize the material distribution within a macroscopic body, e.g., obtaining the fill-fraction field $\rho(\bfx)$ for all $\bfx\in\Omega$ by minimization of the total compliance subject to given boundary conditions and loads.

Multiscale topology optimization additionally focuses on the microscale design (e.g., optimizing the architecture of a metamaterial, or the fiber orientation in composites) in a spatially-variant fashion along with the material distribution on the macroscale (unlike, e.g., multiresolution approaches which gain efficiency by introducing different mesh resolutions on the same (macro-)scale \citep{Nguyen2010}). The two-scale optimization may be carried out sequentially  \citep{schury2012efficient,zowe1997free} -- e.g., identify optimal material stiffness tensor fields for the macroscale problem and then search the microstructural design space to meet the target properties using  inverse homogenization methods \citep{sigmund1994materials,sigmund1995tailoring}. In practice, this approach faces challenges because the target material properties may  be physically infeasible or unachievable by the chosen microstructural design space. As a remedy, simultaneous optimization at both scales (e.g., in an FE$^2$ setting) is more robust \citep{rodrigues2002hierarchical,coelho2008hierarchical,xia2014concurrent} but also computationally expensive. Here, we adopt the latter approach of simultaneous two-scale optimization in a new, computationally efficient fashion.

Topology optimization with anisotropic materials allows for manipulating the material orientation to generate structures with superior properties. In the context of compliance minimization problems, several works have shown that optimal orientations of anisotropic materials tend to align with the principal stress or strain directions \citep{suzuki1991homogenization,gibiansky2018microstructures,diaz1992shape,pedersen1989optimal,pedersen1990bounds,pedersen1991thickness,gao2012bi,stegmann2005discrete,groen_homogenization-based_2019}. Within two-scale topology optimization, such alignment can be achieved by treating, e.g., the orientation angle(s) as additional design variables, which in composites is also known as \textit{continuous fiber angle optimization} (CFAO) \citep{jiang2019cfao,xia2017cfao,setoodeh2005combined,nomura2015general}. In most such approaches, the inherent anisotropy of the base material is assumed constant. By contrast, recent works \citep{sivapuram2016simultaneous, gao2019topology, watts2019simple, white2019multiscale} optimized the material anisotropy by tuning the microstructural architecture; yet, they did not consider the (spatially varying) orientation of the microstructure in a macroscopic body -- primarily because such designs are based on periodic unit cells, which do not admit tessellations with arbitrary and/or spatially-variant orientations. For example, the optimized strut-based cuboidal unit-cells of \citet{watts2019simple} \revised{(while improving manufacturability)} are always aligned with the Cartesian coordinate axes, which is sub-optimal compared to the case when the cells and the constituent struts are aligned locally with the principal stress or strain directions. Recently, \citet{wu2019conforming} introduced \textit{conforming lattice optimization} to address this issue; however, their strut-based design domain (rectangular or cuboidal cells with beams at the edges) strongly limits the microstructural and anisotropic tunability. Distinct from the above works with either fixed material anisotropy or fixed material orientation, our spinodoid topologies discussed here are simultaneously optimized for both the material anisotropy as well as the orientation distribution of the microstructure within a macroscopic body.

When designing metamaterials with spatially varying microstructure, multiscale topology optimization is computationally expensive, as on-the-fly homogenization methods using, e.g., the finite element method (FEM) must extract the effective properties at each material point on the macroscale from the underlying, local microstructure in each iteration of the optimization problem. Look-up tables are a convenient short-cut \citep{SchumacherEtAl2015}, yet they strongly limit the available design space and raise questions about interpolations between available data. More recently, machine learning (ML) techniques have attracted interest for accelerating topology optimization in two ways. First, by creating a dataset of solutions obtained using conventional multiscale topology optimization for several different boundary conditions, loads, microstructures, and material properties, a data-driven model can be learned for accelerated or even real-time prediction of the optimal solution as a function of these inputs without the need for an optimizer \citep{ulu2016data,sosnovik2019neural,banga20183d,lei2019machine,yu2019deep,zhang2019deep}. However,  since these approaches are essentially image-to-image learning (e.g., treating the boundary conditions or material distribution as multi-channel images), a large amount of training data is required to achieve reasonable accuracy. Additionally, generalization to unseen inputs (e.g., new loads or boundary conditions) is limited. The above methods are further challenging to extend to three-dimensional (3D) multiscale optimization with high-dimensional design parametrizations.

Second, topology optimization has been accelerated by employing homogenization surrogate models, which are typically developed through an offline training or interpolation of structure-property tables. For example, \citet{watts2019simple} developed polynomial surrogate models of homogenized elastic properties of open-truss micro-architectures for deployment in multiscale topology optimization. Others have used Gaussian processes \citep{zhang2019concurrent}, Numerical EXplicit Potentials (NEXP) \citep{xia2015multiscale,yvonnet2009numerically}, and single-layer neural networks (NNs) \citep{white2019multiscale} in related contexts to approximate the effective material response to bypass expensive FEM simulations. However, complex design spaces like that of spinodoids (which have high-dimensional, highly-nonlinear, and multiply-connected parametrization domains) require surrogate models beyond simple interpolation methods. To this end, we employ deep NNs (DNNs), which have emerged as powerful tools to efficiently learn in high-dimensional spaces. In this contribution, a DNN pre-trained on a structure-property dataset (obtained via FEM)  is used to predict the anisotropic elastic stiffness of a given spinodoid topology.

In this study, we present a multiscale topology optimization framework for cellular structures based on spinodoid topologies, with simultaneous optimization of  the macroscale material distribution and the microstructural (spinodoid) design and orientation, where the effective microscale response and associated sensitivities are provided by a data-driven surrogate model that replaces nested FE calculations on the microscale. We point out that the optimization problem requires computing the sensitivity of the compliance at a material point on the macroscale with respect to its design parameters. \LZ{Lacking closed-form derivatives, this sensitivity analysis in our multiscale description requires numerical differentiation (ND),} which involves perturbing the design parameters and recomputing the effective stiffness for each perturbation -- requiring computationally expensive FE calculations. Moreover, its accuracy is strongly subject to round-off and truncation errors. While symbolic differentiation (SD) of the surrogate model can give exact derivatives, it leads to inefficient and redundant expressions when applied to complex nonlinear functions as is the case here. We address this issue by using automatic differentiation (AD), which is naturally supported by our NN-based surrogate model. Distinct from ND and SD, AD avoids the above limitations by creating a computational graph that tracks the series of mathematical operations between the inputs and outputs of an arbitrary function and applying chain rule recursively to compute its  derivatives. Note that the derivatives are exact, even though AD does not provide the functional form of the derivatives like SD. An extensive introduction to AD can be found in \citet{griewank2008evaluating}. Examples of application of AD to design optimizations can be found in, e.g., \cite{su1997automatic,barthelemy1995automatic,charpentier2012higher,white2019multiscale}.

In the following, we present the topology generation and homogenization of spinodoids in Section~\ref{sec:spinodoids}. Section~\ref{sec:topOpt} defines the multiscale topology optimization problem, for which Section~\ref{sec:Data-driven surrogate model} introduces the DNN-based surrogate model and discusses its implementation in topology optimization. We use this model in Section \ref{sec:Results} to present several benchmark examples of compliance minimization, before Section~\ref{sec:Conclusions} concludes our study.


\section{Spinodoid topologies with tunable anisotropy}
\label{sec:spinodoids}

\subsection{Spinodoid topology generation}\label{sec:topology_generation}

Consider a homogeneous two-phase solution occupying a domain $\calV\in \Rset^3$ and undergoing spinodal decomposition. During the early stages, the separated phases are spatially-arranged in a stochastic bi-continuous topology, whose evolution is governed by the Cahn-Hilliard equation \citep{cahn1958free,cahn1961spinodal,cahn1965phase}. Let $\varphi:\calV\to\Rset$ be the phase field that describes the concentration fluctuation of one phase. Using Fourier analysis, Cahn showed that the phase field can be described by a GRF --  a superposition of several plane waves with fixed wavelength but random wave vectors and phase shifts. That is, the concentration $\varphi$ of either of the two phases at $\bfx \in \calV$ is given by
\begin{equation}\label{eq:GRF}
	\varphi(\boldface{x}) = \sqrt{\dfrac{2}{N}}\sum\limits_{i = 1}^{N\gg 1} \cos\left(\beta \boldface{n}_i \cdot \boldface{x}+\gamma_i\right),
\end{equation}
where $N$ is the number of waves, $\beta>0$ is a constant wavenumber, and $\boldface{n}_i \sim \calU(S^2)$  and $\gamma_i\sim \calU\left([0,2\pi)\right)$ denote the uniformly-distributed direction\footnote{$S^2=\{\bfk\in\Rset^3 : \|\bfk\| = 1\}$ denotes the three-dimensional sphere of unit radius.} and phase angle of the $i^\text{th}$ wave vector, respectively. Without loss of generality, the amplitudes are assumed equal and constant to ensure that $\varphi$ is normally distributed with zero mean and unitary standard deviation. Note that $\beta$ prescribes a wavelength and hence directly controls the microstructural length scale.

\begin{figure}
	\centering
	\hglue -1.2cm
	\begin{subfigure}{0.4\textwidth}
		\centering
		\includegraphics[width = \textwidth]{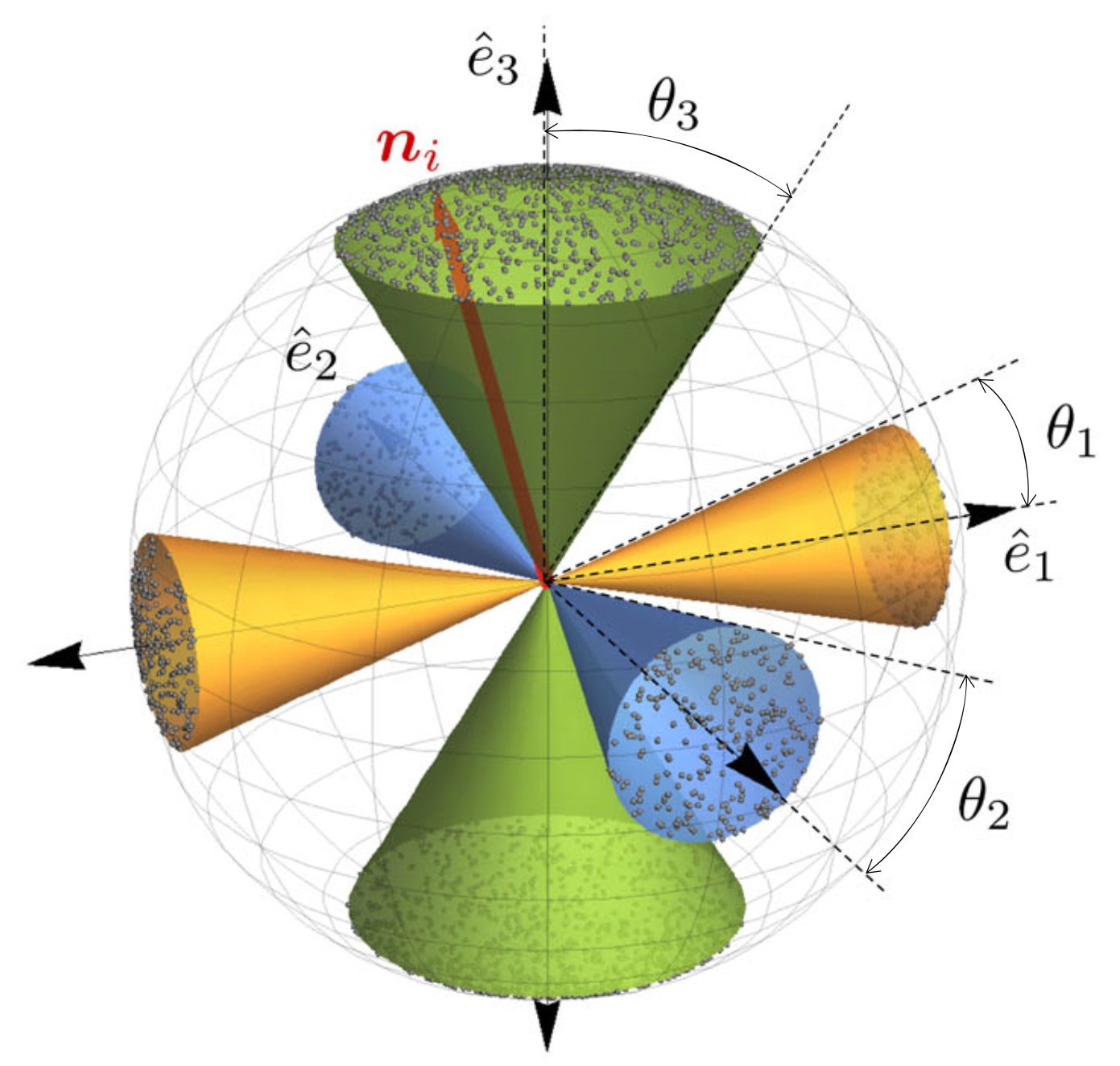}
		\caption{Schematics of the design parameters: $\theta_1, \theta_2, \theta_3$}
		\label{fig:conesWithPoints}
	\end{subfigure}\\

	\centering
	\begin{subfigure}{0.48\textwidth}
		\centering
		\includegraphics[width = \textwidth]{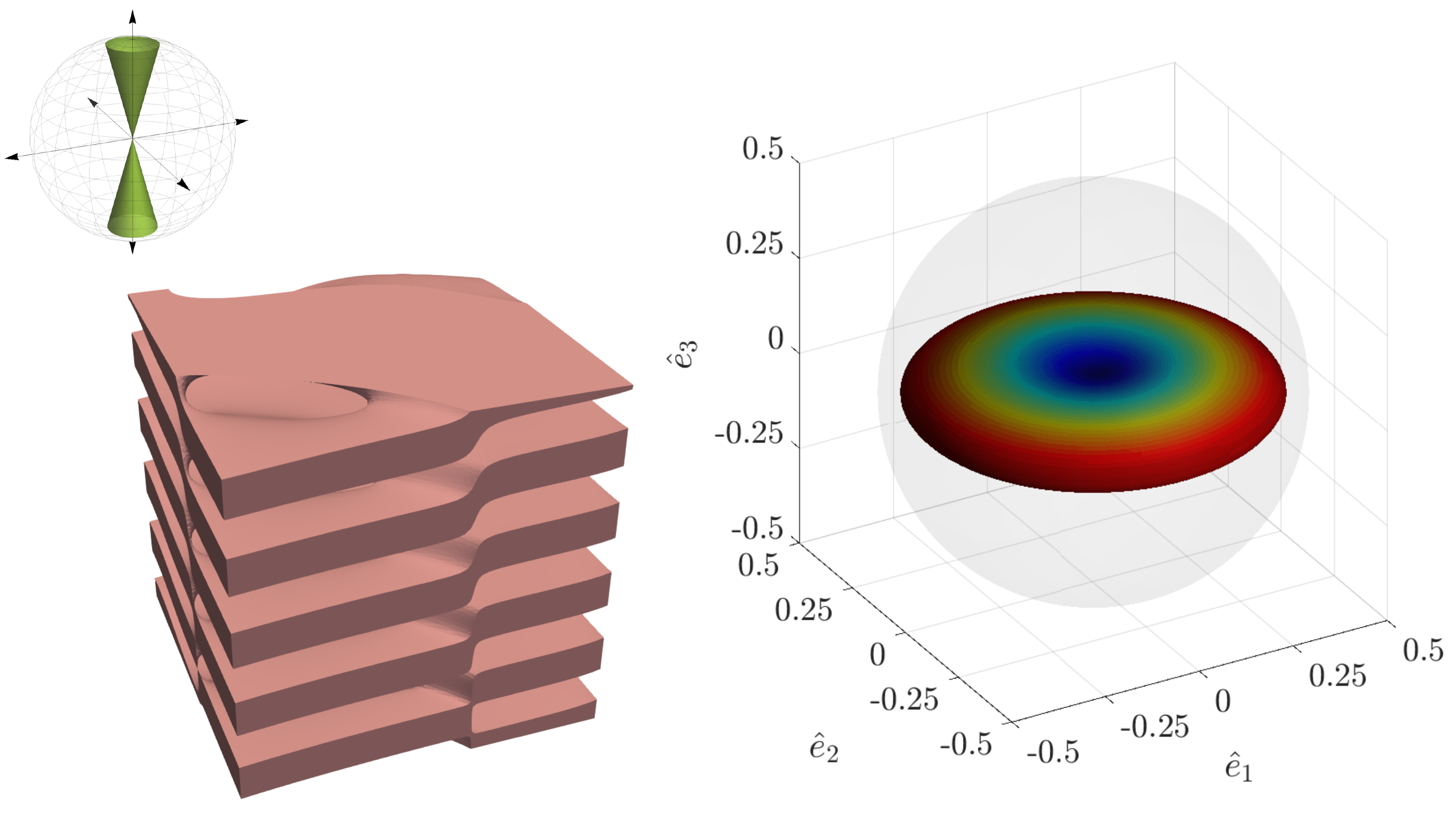}
		\caption{Lamellar: $\rho=0.5, \theta_1 = 0^\circ, \theta_2 = 0^\circ, \theta_3 = 15^\circ$}
		\label{subfig:lamellar}
	\end{subfigure}
	\begin{subfigure}{0.48\textwidth}
		\centering
		\includegraphics[width = \textwidth]{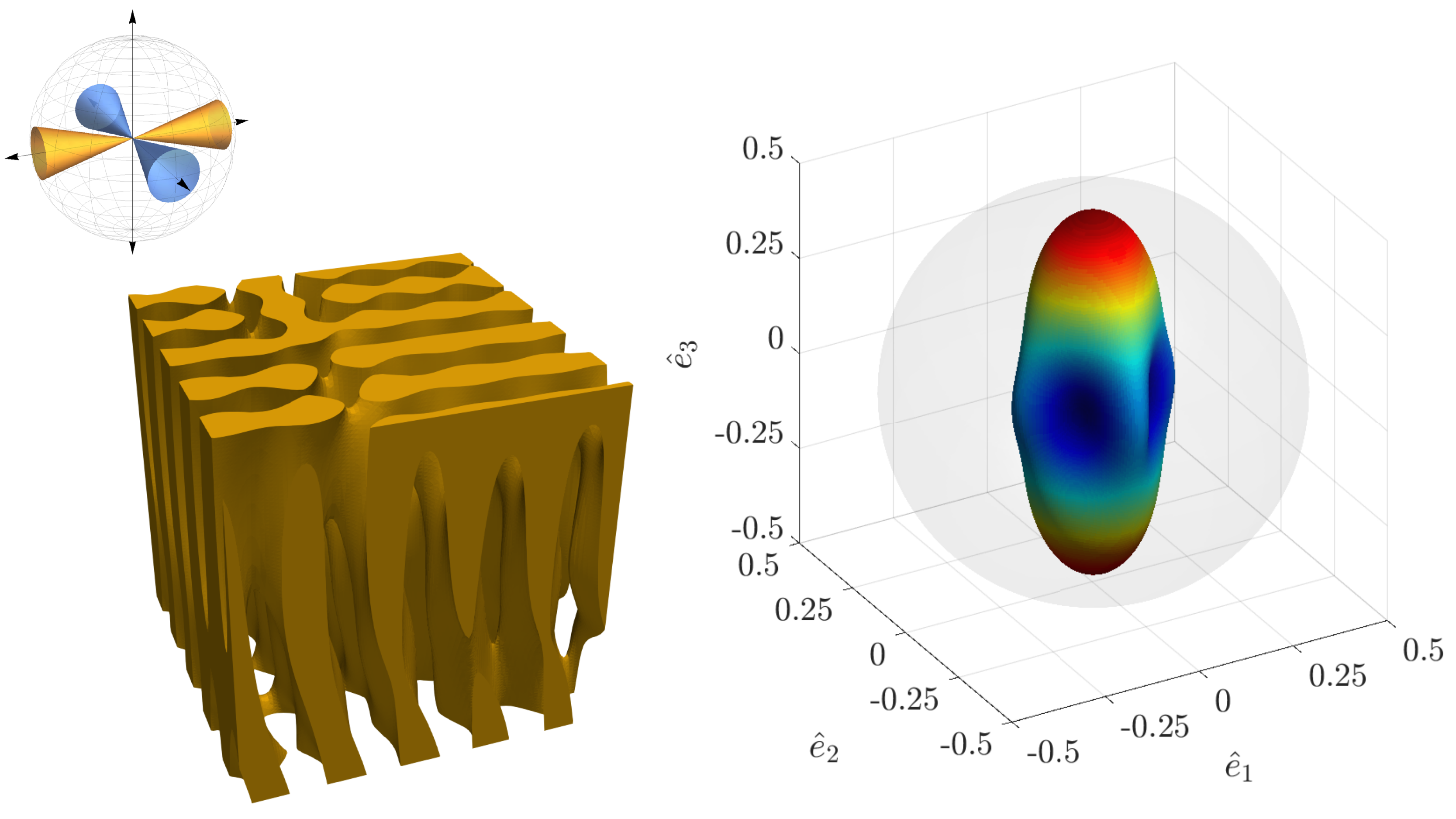}
		\caption{Columnar: $\rho=0.5, \theta_1 = 15^\circ, \theta_2 = 15^\circ, \theta_3 = 0^\circ$}
	\end{subfigure}
	\begin{subfigure}{0.48\textwidth}
		\centering
		\includegraphics[width = \textwidth]{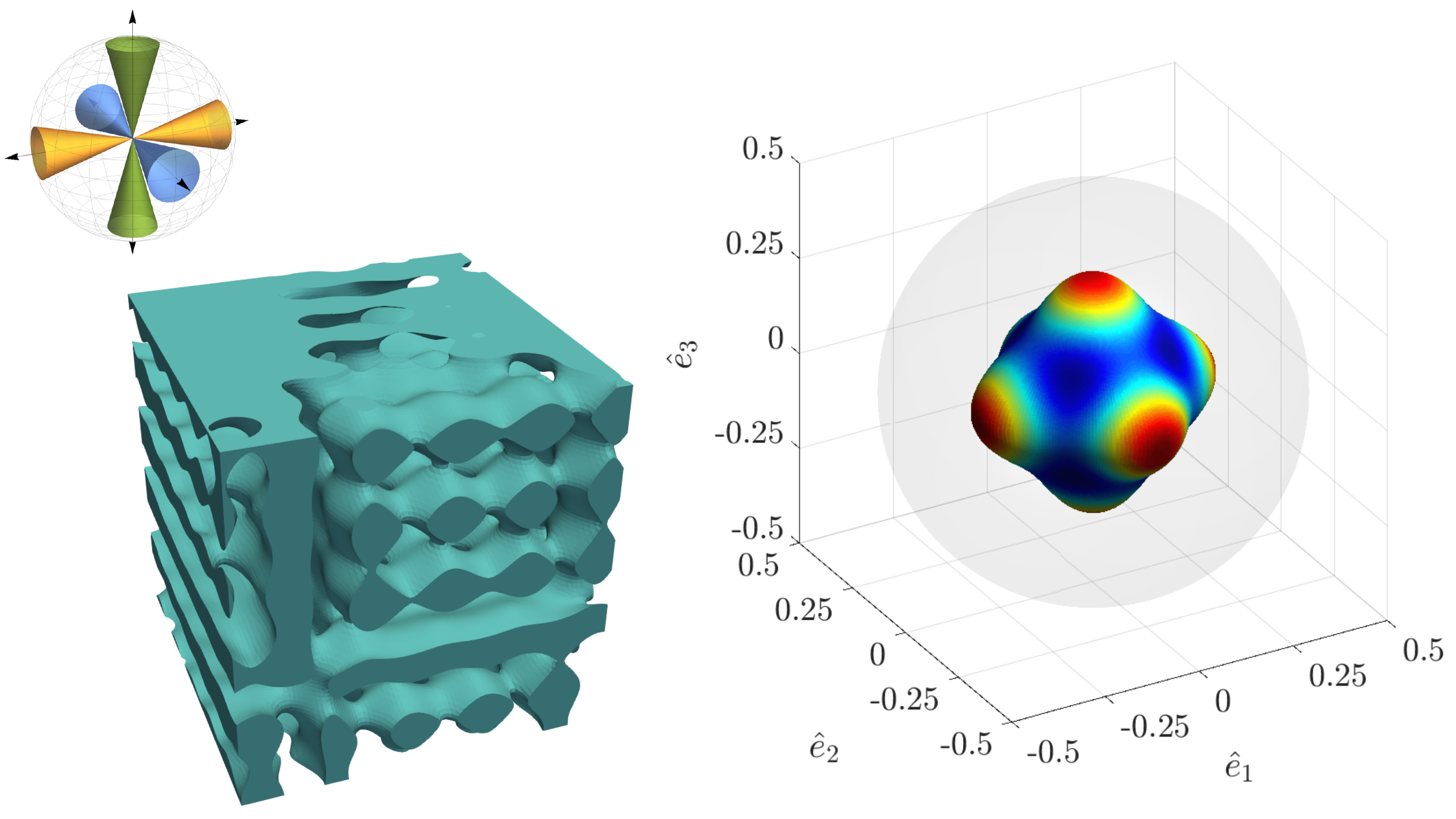}
		\caption{Cubic: $\rho=0.5, \theta_1 = 15^\circ, \theta_2 = 15^\circ, \theta_3 = 15^\circ$}
	\end{subfigure}
	\begin{subfigure}{0.48\textwidth}
		\centering
		\includegraphics[width = \textwidth]{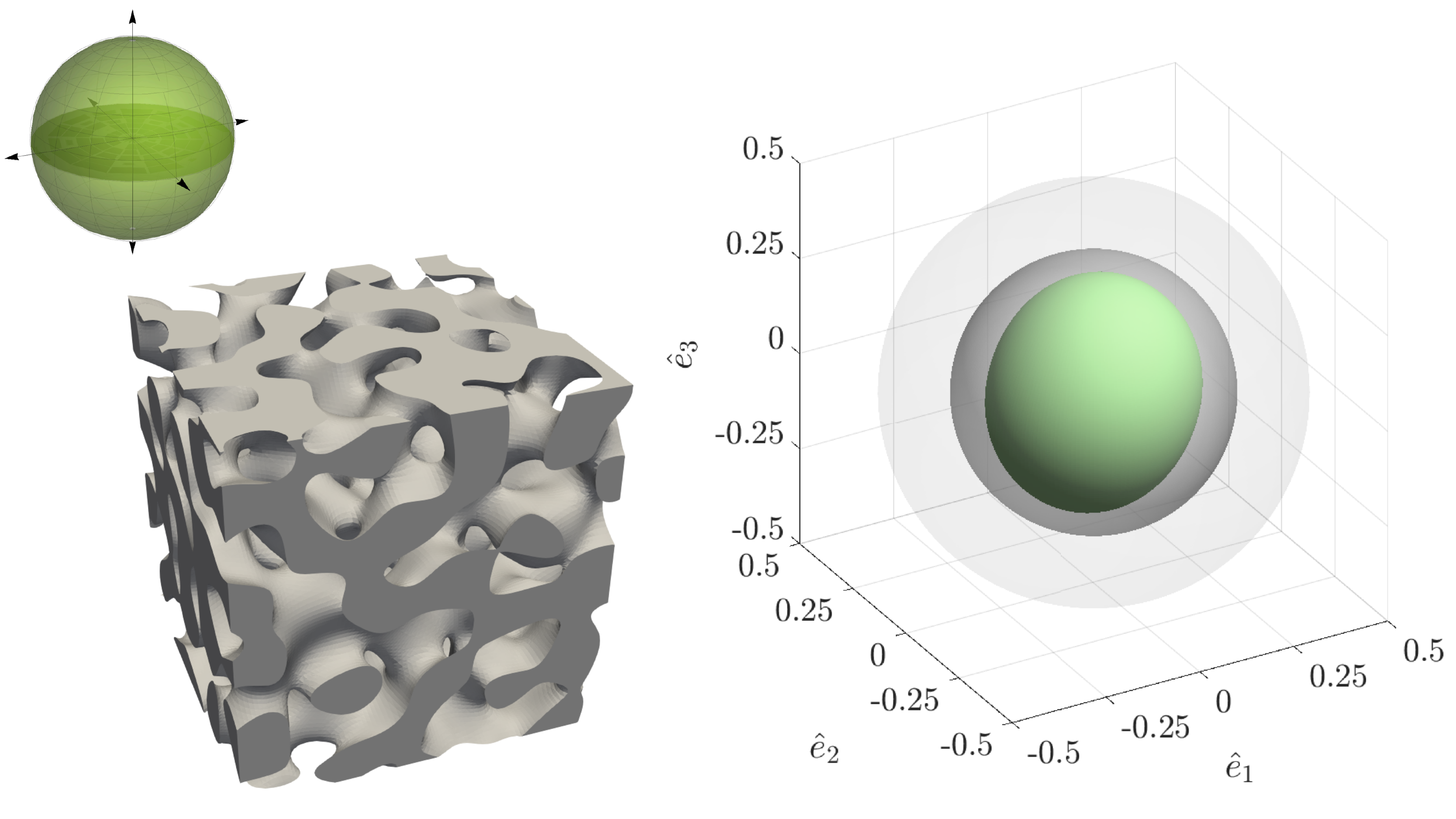}
		\caption{Isotropic: $\rho=0.5, \theta_1 = 90^\circ, \theta_2 = 90^\circ, \theta_3 = 90^\circ$}
	\end{subfigure}
	\caption{Spinodoid topology generation and representative examples. (a) Schematic illustration of the three cone angles $\theta_1$, $\theta_2$, and $\theta_3$, which describe the anisotropic distribution of wave vectors $\bfn_i$. (b-e) Examples of spinodoid topologies with their respective anisotropic elasticity surfaces (and an illustration of the cone angles indicated in each caption). Each point on the elasticity surface denotes the Young's modulus in the corresponding direction. Light and dark gray spheres indicate, respectively, the elastic Voigt upper bound and the Hashin-Shtrikman upper bound (the latter being applicable only to the isotropic structure). Adapted from \cite{kumar2020inverse}. }
	\label{fig:spinodoid topologies}

\end{figure}

The GRF in \eqref{eq:GRF} describes the separated phases in the case of isotropic diffusion, whereas direction-dependent, anisotropic topologies are generally the result phase separation processes with directional preference in interfacial energy or diffusive mobility -- which translates into a directional preference in the distribution of the wave vectors in~\eqref{eq:GRF}. Here, we follow the anisotropic extension of Cahn's GRF solution presented in \citet{kumar2020inverse}. Assuming the principle directions of mobility are aligned with the Cartesian basis $\{\hat{\boldface{e}}_1, \hat{\boldface{e}}_2, \hat{\boldface{e}}_3\}$, the resulting anisotropic topologies can be approximated by a non-uniform orientation distribution function (ODF) parameterized by
\begin{equation}\label{eq:odf}
	\boldface{n}_i \sim \mathcal{U}\left(\left\{\bfk \in S^2 :
	\left(|\bfk\cdot\hat \bfe_1|>\cos\theta_1 \right)\oplus
	\left(|\bfk\cdot\hat \bfe_2|>\cos\theta_2 \right)\oplus
	\left(|\bfk\cdot\hat \bfe_3|>\cos\theta_3 \right)
	\right\}\right),
\end{equation}
such that the wave vectors in \eqref{eq:GRF} are restricted to lie within cone angles $\{\theta_1, \theta_2, \theta_3\}$ from the orthogonal Cartesian basis vectors (see \figurename\ref{fig:conesWithPoints}).

For the purpose of topology generation, GRF models bypass the need for expensive simulations of the Cahn-Hilliard equation. \citet{soyarslan20183d} showed that bi-continuous solid-void microstructures can be generated by applying a level set $\varphi_0$ to the phase field in \eqref{eq:GRF}, such that
\begin{equation}\label{eq:levelset}
\xi(\boldface{x}) = \begin{cases}
1 &\quad  \text{if }\quad  \varphi(\boldface{x}) \leq \varphi_0 \\
0 & \quad \text{if }\quad \varphi(\boldface{x}) > \varphi_0
\end{cases},
\end{equation}
where $\xi(\boldface{x})=\{1,0\}$ denotes the presence of solid and void, respectively, at $\bfx\in\calV$. Since $\varphi$ follows a standard normal distribution, the level set $\varphi_0$ is defined as the quantile evaluated at the average relative density $\rho$ of the solid phase: $\varphi_0 = \sqrt{2}\,\text{erf}^{-1}(2\rho-1)$. We note that, while we only consider solid networks for our multiscale topology optimization framework in the following, shell-type architectures can be generated in a similar fashion by choosing an isosurface of $\varphi(\boldface{x}) = \varphi_0$.

Solid spinodoid topologies are hence characterized by the set of design parameters $\boldface{\Theta} = (\rho, \theta_1, \theta_2, \theta_3)\T$, which provide a way to control the porosity and anisotropy of the architecture. To avoid disjoint solid domains, the design parameters are bounded to $\rho \geq \rho_\text{min}$ and $\theta_1,\theta_2,\theta_3 \in \{0\} \cup [\theta_{\text{min}},\pi/2]$. In subsequent numerical examples we choose $\rho_\text{min} = 0.3$ and $\theta_{\text{min}}=\pi/6$.  The resulting anisotropic topologies can be broadly categorized into lamellar, columnar, and cubic types (\figurename\ref{fig:spinodoid topologies}), and they are obtained when, respectively, one, two, or three of the angles $\{\theta_1, \theta_2, \theta_3\}$ are non-zero.


\subsection{Homogenization of the elastic stiffness}\label{sec:homogenization}

For a given set of design parameters, we use \revised{computational homogenization \citep{huet1990application}} via FEM to compute the effective mechanical stiffness of the corresponding spinodoid topology. We consider a cubic representative volume element (RVE), in which the GRF from \eqref{eq:GRF} with level set \eqref{eq:levelset} is used to generate a spinodoid architecture. The base material of the solid is assumed homogeneous, isotropic, linear elastic with Young's modulus $E_s$ and Poisson's ratio and $\nu_s=0.3$ ($E_s$ is arbitrary as it scales linearly with the effective stiffness). For an RVE of size $l\times l\times l$, the wave number $\beta = 10\pi/l$ was found to be \revised{sufficient for extracting converged homogenized stiffness values from the chosen RVE (larger RVE sizes at fixed $\beta$ yielded only marginal variations in the computed effective stiffness values)}. Six numerical experiments are performed on the RVE, one uniaxial stretch and one simple-shear loading along each of the three principal axes. By applying average strains $\langle\boldface{\varepsilon}\rangle$ to the RVE for each of the six cases, the elastic stiffness tensor $\hat\dsC$ is computed by solving the linear system of equations $\langle\bfsigma\rangle = \hat\dsC\,\langle\boldsymbol\varepsilon\rangle$, where $\langle\bfsigma\rangle=\frac{1}{|\calV|}\int_\calV\bfsigma\dd V$ is the volume-averaged stress obtained from FEM. Since the spinodoid topologies lack periodicity, we apply affine boundary conditions to the RVE and acknowledge that the computed response provides an upper bound to the actual effective stiffness (yet, spinodoids -- like spinodals -- retrieve their beneficial stiffness scaling properties from being stretching-dominated, so that affine boundary conditions can be expected to not affect the response as much as in, e.g., bending-dominated trusses).

The fourth-order stiffness tensor is visualized through the elastic surface, showing Young's modulus $E(\bfd)$ along all directions $\boldface{d}\in S^2$, which is computed as
\begin{equation}
E(\boldface{d}) = \bigg(\sum\limits_{i,j,k,l=1}^3 \hat\dsC_{ijkl}^{-1}d_id_jd_kd_l\bigg)^{-1}.
\end{equation}
\figurename\ref{fig:spinodoid topologies} illustrates representative examples of the diverse anisotropic stiffnesses achievable by the spinodoid design space -- from lamellar topologies (being highly soft in a dedicated direction) to columnar ones (stiff in two directions) to cubic and (approximately) isotropic topologies. In subsequent derivations, the strain, stress, and stiffness tensors will be used frequently in their respective Voigt notations, which we denote by subscript $(\cdot)_v$. For further details on topology generation and homogenization of spinodoids, the reader is referred to \citet{kumar2020inverse}.


\section{Multiscale topology optimization}\label{sec:topOpt}

\begin{figure}
	\centering
	\includegraphics[width =0.65 \textwidth]{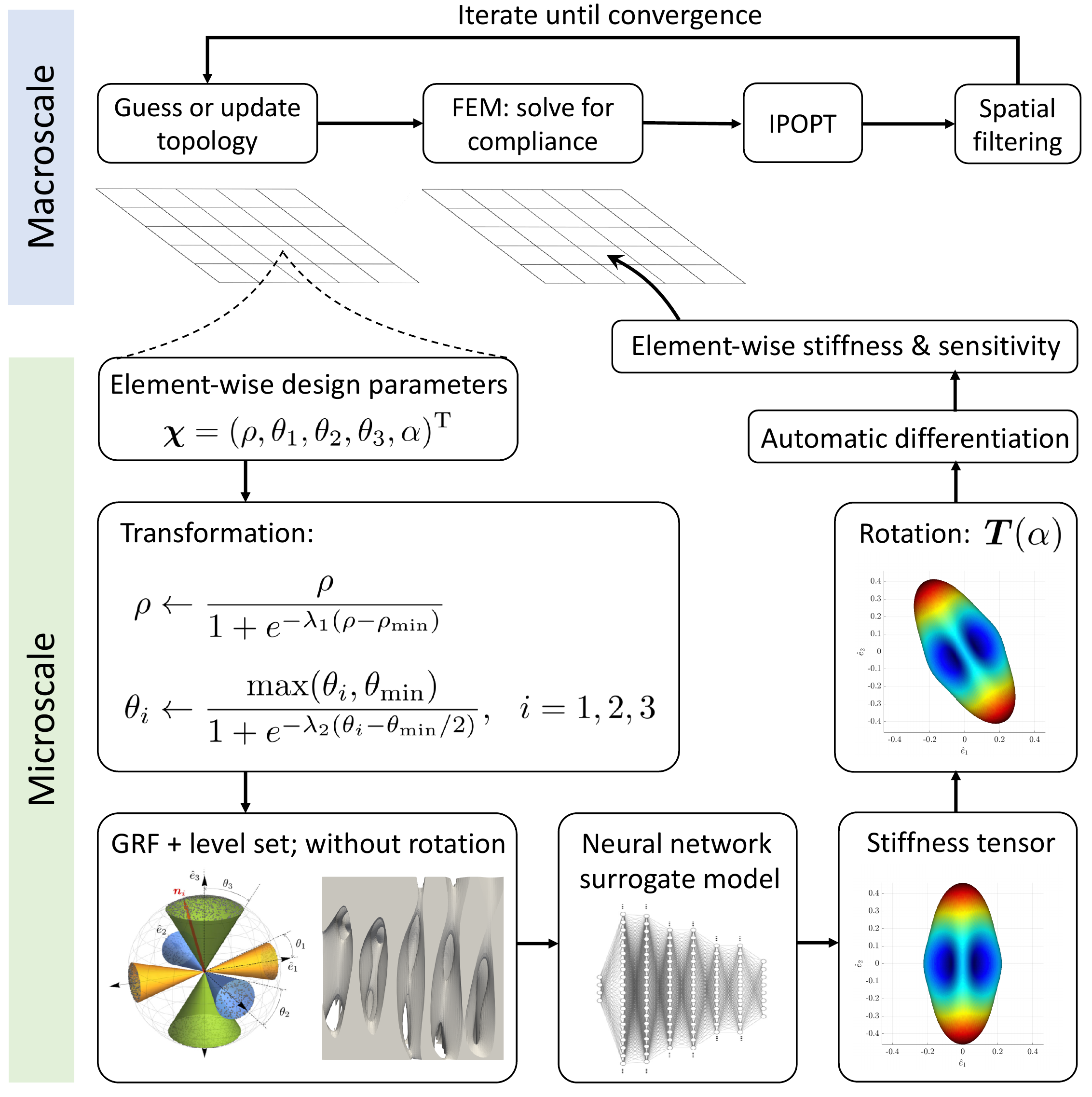}
	\caption{Schematic illustration of the topology optimization framework for spinodoid metamaterials.}\label{fig:flowChart}
\end{figure}

The overall multiscale topology optimization setup presented in the following is schematically summarized in \figurename\ref{fig:flowChart}.

\subsection{Macroscale: boundary value problem and compliance minimization}\label{sec:macroscale}

The classical boundary value problem (BVP) of static equilibrium in a macroscale domain $\Omega\subset\Rset^d$ in $d$ dimensions is given by
\be\label{eq:bvp}
	\divv \bfsigma(\bfu,\bfchi) = \bfzero \ \ \   \text{in} \  \Omega,\qquad
	\bfu = \tilde\bfu\ \ \  \text{on}\ \partial\Omega_u,\qquad
	\bfsigma \bfn = \tilde\bft\ \ \  \text{on}\ \partial \Omega \setminus \partial \Omega_u.
\ee
Here, $\bfu:\Omega\to\Rset^d$ denotes the displacement field with essential boundary conditions $\tilde \bfu$ on the Dirichlet boundary $\partial\Omega_u\subseteq \partial \Omega$. $\tilde\bft$ denotes the prescribed tractions on the remaining boundary with outward unit normal $\bfn$, and $\bfsigma$ is the symmetric infinitesimal stress tensor. For linear elasticity with the linearized strain tensor $\bfeps=\sym(\nabla\bfu)$, stresses and strains are locally related via $\bfsigma(\bfx)=\dsC(\bfx)\bfeps(\bfx)$ for $\bfx\in\Omega$.

In our two-scale setup, we assume that the local elastic properties and mass density at a point $\bfx\in\Omega$ on the macroscale stem from homogenization over a microscale RVE, whose architecture is defined by a local set of design parameters, $\bfchi(\bfx)$. Assuming a separation of scales, the macroscale stresses $\bfsigma$ and strains $\bfepsilon$ are related via the linear elastic constitutive relation
\be\label{eq:constitutive_law}
	\bfsigma(\bfx)=\mathbb{\hat C}\big(\bfchi(\bfx)\big)\bfepsilon(\bfx),
\ee
where $\mathbb{\hat C}$ represents the homogenized stiffness tensor obtained from an RVE with design parameters $\bfchi$, so the spatial variability in microstructural topology is captured naturally by $\bfchi(\bfx)$. Macro- and microscales are hence coupled by exchanging homogenized information such as $\bfchi$ and $\hat\dsC$. For the example of spinodoids, $\bfchi$ includes $\bfTheta = (\rho,\theta_1,\theta_2,\theta_3)^T$ among other design parameters which will be explained in Section~\ref{sec:microscale}.

We use FEM to solve the macroscale BVP \eqref{eq:bvp} in a (Bubnov-)Galerkin setting, which discretizes the macroscale body $\Omega$ into a mesh containing $n$ elements and $n_n$ nodes. In subsequent examples, we use constant-strain tetrahedra in 3D, having a single quadrature point per element. Of course, the following formulation can be analogously adopted for higher-order elements. Let $\calX = \{\bfchi^e, e=1,\dots,n\}$ be the set of all microstructural design parameters  (one such set for each element $e$), such that the stiffness matrix of element $\Omega^e$ is given by
\be\label{eq:k_elem}
	\bfk^e(\bfchi^e) = \int_{\Omega^e}({\bfB^e})\T \mathbb{\hat C}_v(\bfchi^e) \bfB^e \dd \Omega,
\ee
where $\bfB^e$ denotes the strain-displacement matrix of element~$e$ \citep{CookEtAl2001}. For given $\calX$, the global vector of nodal displacements $\bfU=\{\bfu_i, i=1,\ldots,n_n\}$ is obtained from the following system of equations (assuming that essential boundary conditions have been imposed):
\be\label{eq:KU=F}
	\bfK(\calX)  \bfU(\calX) =\bfF.
\ee
The global displacement vector $\bfU$, global stiffness matrix $\bfK$, and the external force vector $\bfF$ are obtained via assembly of the element-wise displacement vectors $\bfu^e$, element stiffness matrices $\bfk^e$, and external forces (derived for each element from the applied tractions $\tilde \bft$), respectively.

The total compliance at the macroscale is given by
\be\label{eq:compliance}
 \Phi(\calX)
	= \boldface{U}(\calX)\cdot\boldface{F}
	= \boldface{U}(\calX)\cdot\boldface{K}(\calX)\boldface{U}(\calX)
	= \sum_{e=1}^n {\bfu^e}\cdot\bfk^e \bfu^e
	= \sum_{e=1}^n {\bfu^e}\cdot\left( \int_{\Omega^e}(\bfB^e)\T \mathbb{\hat C}_v(\bfchi^e) \bfB^e \dd \Omega\right) \bfu^e,
\ee
where \eqref{eq:k_elem} and \eqref{eq:KU=F} have been substituted. The compliance minimization problem on the macroscale is formulated as
\be\label{eq:objective}
       \min_\calX \Phi(\calX), \quad \text{subject to} \quad  \frac{1}{|\Omega|} \int_\Omega\rho \dd V\leq \bar\rho \quad \text{and}\quad  \text{\eqref{eq:KU=F}},
\ee
where $\bar\rho$ prescribes a target on the overall relative density at the macroscale (effectively imposing a total weight constraint), and \text{\eqref{eq:KU=F}} ensures static equilibrium of the loaded structure.

The nonlinear optimization problem in \eqref{eq:objective} is solved using IPOPT \citep{wachter2006implementation}, a primal-dual interior-point algorithm with a filter line-search method. IPOPT requires sensitivity information, i.e., the derivative of the compliance $\Phi$ with respect to the design parameters $\calX$, for determining the line search direction. Differentiating both sides of the equilibrium constraint \eqref{eq:KU=F} with respect to $\calX$ and noting that $\bfK$ is symmetric, we obtain
\be\label{eq:constraint_deriv}
	\partderiv{\bfK(\calX)\bfU(\calX)}{\calX}  = 0
	\qquad \Rightarrow \qquad \bfK\partderiv{\bfU}{\calX}
	= - \partderiv{\bfK}{\calX}\bfU \quad \text{and} \quad \partderiv{\bfU}{\calX} \cdot\bfK
	= -\bfU\cdot\partderiv{\bfK}{\calX}.
\ee
The compliance sensitivity hence follows as
\be
\partderiv{\Phi}{\calX}  = \partderiv{\bfU\cdot \bfK \bfU}{\calX} = -\bfU\cdot\partderiv{\bfK}{\calX}\bfU,
\ee
where \eqref{eq:constraint_deriv} has been substituted to obtain the right-hand side. Using \eqref{eq:compliance}, the above may be expressed as an element-wise summation, viz.
\begin{equation}\label{eq:compliance_sensitivity}
	\dfrac{\partial \Phi}{\partial \calX}
	= -\sum_{e=1}^n {\bfu^e}\cdot \left(\int_{\Omega^{e}} (\bfB^e)\T  \partderiv{\mathbb{\hat C}_v(\bfchi^e)}{\bfchi^e} \bfB^e \dd\Omega\right)\bfu^e,
\end{equation}
where $ \partial  \mathbb{\hat C}_v(\bfchi^e) /\partial\bfchi^e$ is the element-wise or microscale stiffness sensitivity.

To eliminate numerical instabilities such as checkerboard patterns, we impose a minimum length scale by a filtering technique. Commonly used filter techniques include density filters \cite{bruns2001topology}, sensitivity filters \cite{sigmund1997design}, and morphology-based filters \cite{sigmund2007morphology}. In our approach, the stiffness sensitivity in \eqref{eq:compliance_sensitivity} is replaced by (no summation over $i$ implied)
\be\label{eq:filter}
	\partderiv{\mathbb{\hat C}_v(\bfchi^e)}{\chi^e_i}
	\leftarrow \left(\displaystyle\sum_{e'=1}^n w\left(\bfx,\bfx^{e'}\right)  \chi^{e'}_i \partderiv{\mathbb{\hat C}_v(\bfchi^{e'})}{\chi^{e'}_i}\right)
	\left(\chi^{e}_i\displaystyle \sum_{e'=1}^n w\left(\bfx,\bfx^{e'}\right)  \right)^{-1}
\ee
with weight function
\be
	w\left(\bfx,\bfx^{e'}\right)
	= \max \left( 0, r_\text{filter} - \left\| \bfx^e - \bfx^{e'} \right\|\right),
\ee
where $\bfx \in \Omega^e$, while $\bfx^e$ denotes the position of the center of element $e$, and $r_\text{filter}>0$ is a cut-off radius that controls the length scale of the filter.

The optimal solution of \eqref{eq:objective} yields the set of spatially-varying design parameters $\calX$, which can be used to generate the multiscale topology with fully resolved microstructures. Details on the computational generation of fully resolved topologies are presented in \ref{sec:spatiallyVariantAppendix}.


\subsection{Microscale: spinodoid microstructures}
\label{sec:microscale}

The spinodoid microstructure at every point $\bfx\in\Omega$ on the macroscale depends on design parameters $\bfchi(\bfx)$. As we are using simplicial elements (with a single quadrature point), we deal with element-wise constant $\bfchi^e$. Here, we consider the spinodoid topology within a single element and, for the sake of brevity, we drop the superscript $(\cdot)^e$ in the following discussion, while tacitly implying that the microscale description applies for each element $e\in \{1,\dots,n\}$. For spinodoids, the microscale response is generally determined by design parameters $\bfchi= \bfTheta = (\rho,\theta_1,\theta_2,\theta_3)\T$. Unfortunately, this straightforward definition of $\bfchi$ presents technical challenges in the context of multiscale topology optimization, which are addressed here.

\begin{figure}[t]
	\centering
	\begin{subfigure}{0.4\textwidth}
		\includegraphics[width = \textwidth]{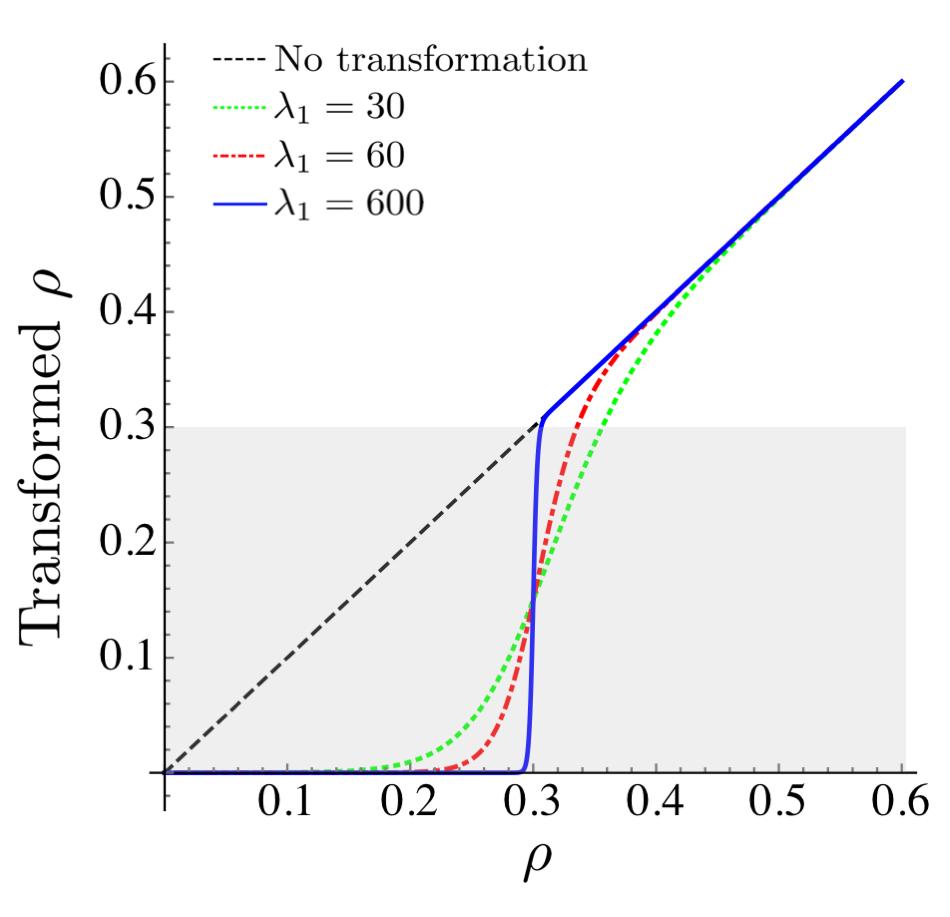}
		\caption{}\label{fig:transform1}
	\end{subfigure}
	$\qquad$
	\begin{subfigure}{0.4\textwidth}
		\includegraphics[width = \textwidth]{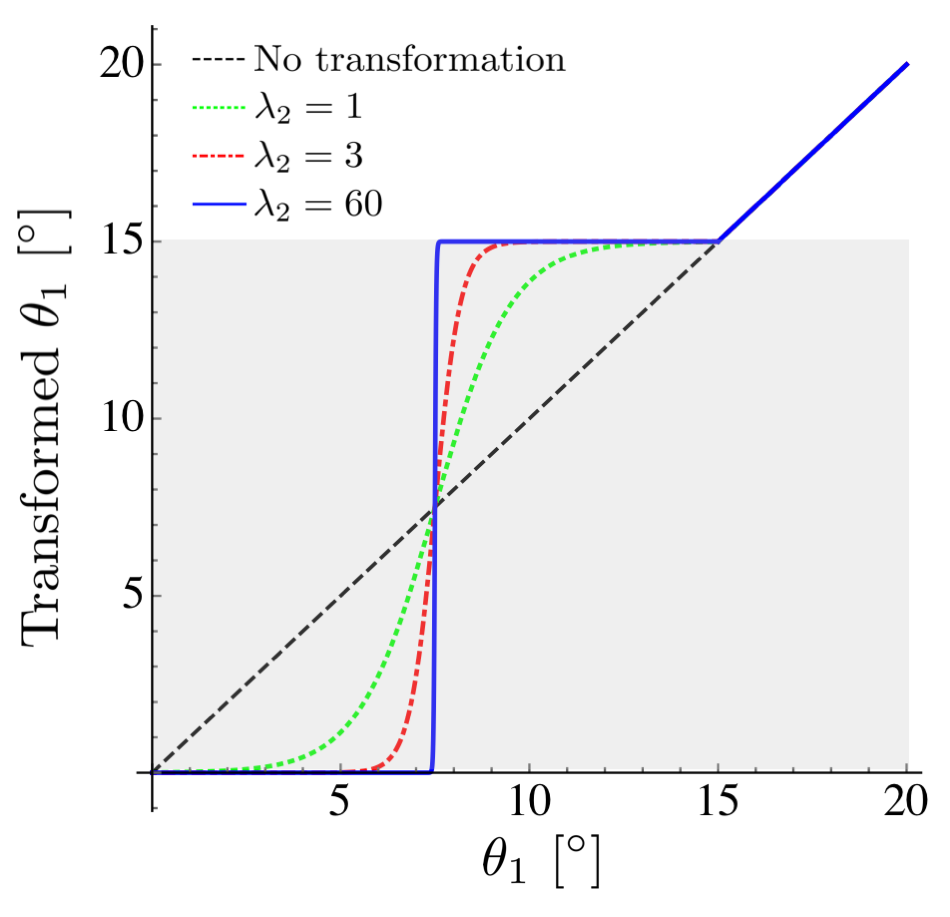}
		\caption{}\label{fig:transform2}
	\end{subfigure}
\caption{Transformation of the design parameters (a)  $\rho$ and (b) $\theta_1$ according to $\bff(\bfTheta)$ in \eqref{eq:transform} for different values of $\lambda_1$ and $\lambda_2$, introduced to avoid inadmissible values (indicated by the shaded regions corresponding to (a) $\rho_{\min}=0.3$, (b) $\theta_{\min}=15^\circ$).}\label{fig:transform}
\vskip 0.3cm
	\centering
		\includegraphics[width = \textwidth]{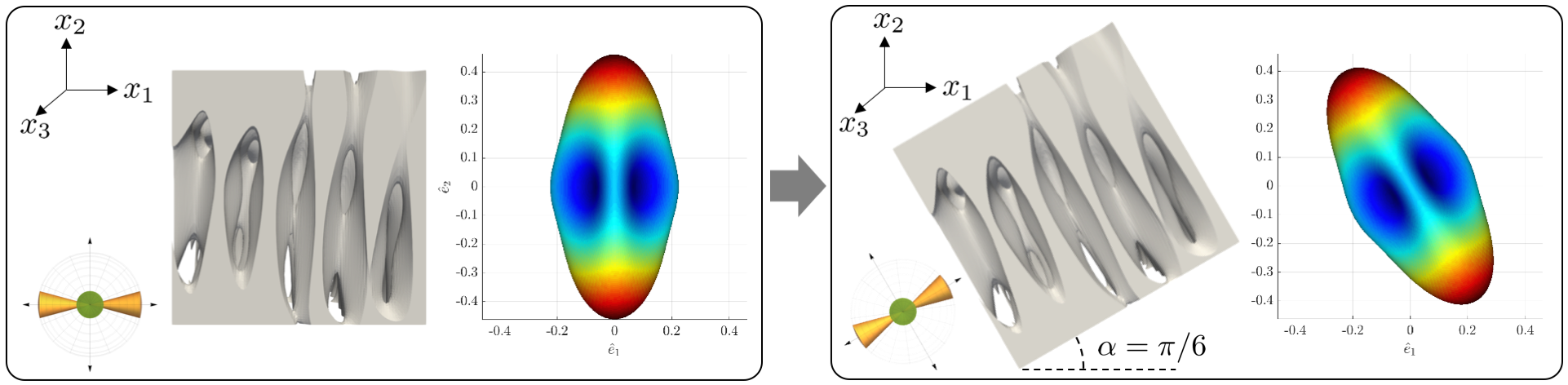}
	\caption{Rotation (with $\alpha=30^\circ$) of a representative spinodoid topology ($\rho=0.5,\theta_1=15^\circ, \theta_1=0^\circ,\theta_3=15^\circ$) and its elasticity surface.}\label{fig:rotation}
\end{figure}

As discussed in Section~\ref{sec:spinodoids}, the spinodoid design space is restricted to $\rho\geq\rho_\text{min}$ to avoid disjoint solid domains (at the microscale). With this constraint, however, the entire macroscale domain is filled with material ($\rho>0$). As a remedy, we need to introduce macroscale holes by allowing $\rho=0$ in the (macroscale) voids\footnote{Here, \textit{voids} refer to the absence of material in particular elements in the conventional topology optimization sense, i.e., on the macroscale. This is not to be confused with the bi-continuous topology of solid-\textit{void} phases in the spinodoid architectures on the microscale.}, while enforcing $\rho\geq\rho_\text{min}$ in the non-void regions. \revised{A similar transformation was proposed by \cite{white2019multiscale}}. Similar to the density $\rho$, the domain for angles $\theta_1,\theta_2,\theta_3 \in \{0\}\cup[\theta_{\text{min}},\pi/2]$ is also disjoint. From an optimization perspective, such disjoint parameter spaces are not favorable. Therefore, we transform the parameter space in a continuous fashion by defining the transformation (\figurename\ref{fig:transform})
\be\label{eq:transform}
	\bff(\bfTheta) = \left(
	\dfrac{\rho}{1+e^{-\lambda_1(\rho-\rho_\text{min})}},
	\dfrac{\max(\theta_1,\theta_{\text{min}})}{1+e^{-\lambda_2(\theta_1-\theta_\text{min}/2)}},
	\dfrac{\max(\theta_2,\theta_{\text{min}})}{1+e^{-\lambda_2(\theta_2-\theta_\text{min}/2)}},
	\dfrac{\max(\theta_3,\theta_{\text{min}})}{1+e^{-\lambda_2(\theta_3-\theta_\text{min}/2)}}
	\right)\T \qquad \text{with} \quad \lambda_1,\lambda_2\gg 1.
\ee
Here, each transformation is a smooth approximation of a Heaviside step function \revised{(replacing the discontinuous design space by a $C^0$-continuous function)}, which \revised{allows for} continuous parameter spaces $\rho\in[0,1]$ and $\theta_1,\theta_2,\theta_3\in [0,\pi/2]$ when replacing $\bfTheta$ by $\bff(\bfTheta)$ in the topology optimization problem. The smoothness, controlled by $\lambda_1$ and $\lambda_2$, is necessary to ensure that gradient-based optimization methods such as IPOPT are applicable.  The disallowed parameter values are approximately mapped to the closest allowed values. For example, using $\bff(\bfTheta)$ relaxes the bounds on the relative density to $\rho\in[0,1]$, as microstructures with $\rho<\rho_\text{min}$ are effectively mapped to void with zero stiffness. When weight or compliance minimization is the objective, this will force the density to be either zero (void) or greater than $\rho_\text{min}$ (spinodoid).

So far, the principal directions of the spinodoid topologies and their inherent mechanical anisotropy are aligned with the fixed Cartesian basis $\{\hat\bfe_1,\hat\bfe_2,\hat\bfe_3\}$. In this case, the effective stiffness $\hat\dsC$ in \eqref{eq:constitutive_law} is the homogenized stiffness of the microstructure (Section~\ref{sec:homogenization}). As the stiffness is direction dependent, we here further expand the design space by applying rigid-body rotations to the generated spinodoid topologies. For demonstration purposes (and since our subsequent benchmarks are in 2D), we restrict ourselves to rotations about a single axis, while noting that the framework can be extended to general 3D rotations. We introduce an expanded set of design parameters $\bfchi=(\rho,\theta_1,\theta_2,\theta_3,\alpha)^T$, where $\alpha \in [-\pi/2,\pi/2]$ denotes the angle of a rigid-body rotation (\figurename \ref{fig:rotation}) about the (out-of-plane) $\hat\bfe_3$-axis  (due to the two-fold symmetry, rotations with $\alpha$ and $(\pi-\alpha)$ are equivalent). Applying such a rotation transforms the Voigt-notation stiffness matrix according to
\be
	\mathbb{\hat C}_v(\bfchi) = \bfT(\alpha)\mathbb{C}_v\left(\bff(\bfTheta)\right)\bfT\T(\alpha),
\ee
where $\mathbb{\hat C}_v$ is the effective stiffness passed to the macroscale in \eqref{eq:constitutive_law}, and $\mathbb{C}_v\left(\bff(\bfTheta)\right)$ is obtained from microscale homogenization of the spinodoid topology with design parameters $\bff(\bfTheta)$. The components of rotation matrix $\bfT$ are given by
\be
	\bfT(\alpha) = \begin{bmatrix}
		\cos^2\alpha & \sin^2\alpha & 0 & 0 & 0 & 2\sin\alpha\cos\alpha\\
		\sin^2\alpha & \cos^2\alpha & 0 & 0 & 0 & -2\sin\alpha\cos\alpha\\
		0 & 0 & 1 & 0 & 0 & 0\\
		0 & 0 & 0 & \cos\alpha & -\sin\alpha & 0\\
		0 & 0 & 0 & \sin\alpha & \cos\alpha& 0\\
		-\cos\alpha\sin\alpha & \cos\alpha\sin\alpha & 0 & 0 & 0 & \cos^2\alpha - \sin^2\alpha
	\end{bmatrix}.
\ee

The homogenized stiffness and its sensitivity, which involves computing  $\partial\mathbb{C}_v(\bff(\bfTheta))/\partial \bfTheta$, are the computational bottlenecks (often computed by ND), especially since IPOPT is an iterative algorithm which requires computing the sensitivity of each element several times during the optimization. This motivates the use of data-driven surrogate models that can accelerate the optimization process.

\subsection{\revised{Micro-to-macro transition}}

\revised{
As in classical homogenization, we assume a separation of scales between the spinodoid architectures on the microscale and the characteristic lengths over which fields of interest vary in the macroscale boundary value problem. The spinodoid architectures introduced above are in general scale-independent (as long as no material-level size effects emerge), meaning that the size of the spinodoid features is irrelevant in the homogenization framework and can be chosen arbitrarily a posteriori when generating the spatially variant structures. Specifically, relative density and effective elastic properties are independent of the specific length scale of the spinodal features, which is governed by parameter $\beta$ (the characteristic wavelength of the spinodal microstructures is $2\pi/\beta$). Once an optimal two-scale topology has been identified  in any of the subsequent benchmark problems, the spatially-variant structure can be realized with, in principle, arbitrary values of $\beta$, resulting in finer or coarser microscale features, as needed (and as limited by fabrication constraints). We point out that for spinodoid topologies whose features are not orders of magnitude smaller than the macroscale features, the chosen scheme, strictly speaking, presents an abuse of the homogenization assumption since a separation of scales may break down. Hence, for any given application, the choice of $\beta$ should be a compromise between accuracy, efficiency, and manufacturability of the optimized structures.}


\section{Data-driven surrogate model}\label{sec:Data-driven surrogate model}

\subsection{Bypassing FEM-based homogenization}

To bypass the computational expense of repetitive FEM simulations required for the microscale homogenization, we use a DNN-based surrogate model for the on-demand prediction of the homogenized stiffness tensor $\hat\dsC$. Due to the symmetries in the ODF defined in \eqref{eq:odf}, the anisotropic stiffness tensor $\dsC$ is orthotropic with nine independent elastic moduli, which can be encoded into \revised{$\bfS = (\hat\dsC_{1111}, \hat\dsC_{1122},\hat\dsC_{1133},\hat\dsC_{2222},\hat\dsC_{2233}, \hat\dsC_{3333}, \hat\dsC_{2323},\hat\dsC_{3131},\hat\dsC_{1212})\T$}. The DNN can be interpreted as a composition of several linear and nonlinear transformations that provide a map $\calF_\omega:\bfTheta\rightarrow \bfS$ from the spinodoid design parameters $\bfTheta$ to the compressed representation of the stiffness $\dsC$.

We use the DNN architecture previously developed by \citet{kumar2020inverse}, which translates to the composite transformation
\be\label{eq:dnn}
	\calF_\omega[\bfTheta] =
	\calL_{\omega_7}^{32\rightarrow 9} \circ\mathcal{R}
	\circ \calL_{\omega_6}^{32\rightarrow 32} \circ\mathcal{R}
	\circ \calL_{\omega_5}^{64\rightarrow 32} \circ\mathcal{R}
	\circ\calL_{\omega_4}^{64\rightarrow 64} \circ\mathcal{R}
	\circ\calL_{\omega_3}^{128\rightarrow 64} \circ\mathcal{R}
	\circ\calL_{\omega_2}^{128\rightarrow 128} \circ\mathcal{R}
	\circ\calL_{\omega_1}^{4\rightarrow 128} [\bfTheta].
\ee
Here, $\calL_{\omega_k}^{i\rightarrow j}$ denotes the $k^\text{th}$ linear layer parametrized by the set of weights and biases $\omega_k=\{\bfA_k,\bfb_k\}$ -- and we write $\omega = \{ \omega_i, i=1,\dots,7\}$ -- such that any $\bfz\in \Rset^i$ is transformed according to
\be
	\calL_{\omega_k}^{i\rightarrow j} [\bfz]
	= \bfA_k \bfz + \bfb_k, \quad \text{with} \quad \bfA_k\in\Rset^{j\times i}, \bfb_k\in\Rset^{j}.
\ee
$\mathcal{R}(\cdot) = \max(0,\cdot)$ is the rectified linear activation unit (ReLU) that acts element-wise on the input and introduces nonlinearity to the series of linear transformations. The design parameters $\bfTheta$ are mapped into intermediate high-dimensional spaces, where the surrogate model can efficiently and accurately learn the nonlinear relationship to the homogenized stiffness values $\bfS$. This is particularly advantageous as the stiffness space is highly disjoint because of the discontinuities in the design space (recall that $\theta_1,\theta_2,\theta_3 \in \{0\} \cup [\theta_{\text{min}},\pi/2]$); e.g., the effective stiffnesses of the lamellar and columnar topologies are quite different (cf.~\figurename\ref{fig:spinodoid topologies}).

The above surrogate model is trained on a dataset $\left\{(\bfTheta^i,\bfS^i), i=1,\dots,n_\text{train}\right\}$ containing
$n_\text{train}$ pairs of randomly sampled spinodoid design parameters $\bfTheta$ and their corresponding homogenized stiffnesses $\bfS$ (obtained via FEM). The optimal DNN parameters are obtained using back-propagation and the Adaptive Moment Estimation (Adam) \citep{kingma2014adam} to minimize the mean squared prediction error over the training data
\be
	\omega \leftarrow \arg \min_{\omega} \frac{1}{n_\text{train}}\sum_{i=1}^{n_\text{train}}\|\calF_\omega[\bfTheta^i] -\bfS^i \|^2.
\ee
Once trained, the DNN-based surrogate provides accurate stiffness predictions instantly compared to several minutes for FEM-based homogenization (for the meshes used in our study). As the focus here is on its integration into the multiscale topology optimization framework, we refer the interested reader to \cite{kumar2020inverse} for further details pertaining to the architecture, training, and dataset of the DNN surrogate model.

\subsection{Bypassing ND for sensitivity computations}

The DNN-based surrogate model beneficially provides the sensitivity of the stiffness with respect to the design parameters -- an essential information for the multiscale topology optimization. Conventional approaches based on ND that perturb the design parameters are computationally expensive (requiring extensive FEM-based homogenization simulations) and prone to stability and precision issues. In contrast, our DNN-based surrogate model admits inexpensive computations of the sensitivity information, i.e., $\calF_\omega'[\bfTheta] = \partial \bfS / \partial \bfTheta$, which can be rearranged tensorially into $\partial \hat\dsC_v / \partial \bfTheta$. Since  each transformation in the DNN  \eqref{eq:dnn} is differentiable almost everywhere\footnote{$\mathcal{R}(z)$ is differentiable everywhere except at $z=0$. However, in the context of deep learning and numerical precision, the probability of an input $z$ being exactly zero is close to zero. Therefore, it is a reasonable and well-adopted practice to approximate the derivative of ReLU at $z=0$ with one of its subgradients, which lie in the interval $[0,1]$ -- conventionally chosen to be $R'(0)=0$.}, the sensitivity is easily computed by propagating the gradients from $\bfS$ to $\bfTheta$ via the chain rule for differentiation. From an implementation perspective, this is efficiently achieved by AD, which tracks the computational graph from $\bfTheta$ and $\bfS$. The back-propagation of gradients via AD is implemented in most contemporary software for NNs, as it forms the backbone for training NNs using gradient-based optimization methods such as Adam. Note that the gradients computed by AD are exact and hence do not experience stability or precision issues.


\begin{figure}[!t]
	\centering
	\includegraphics[width = \textwidth]{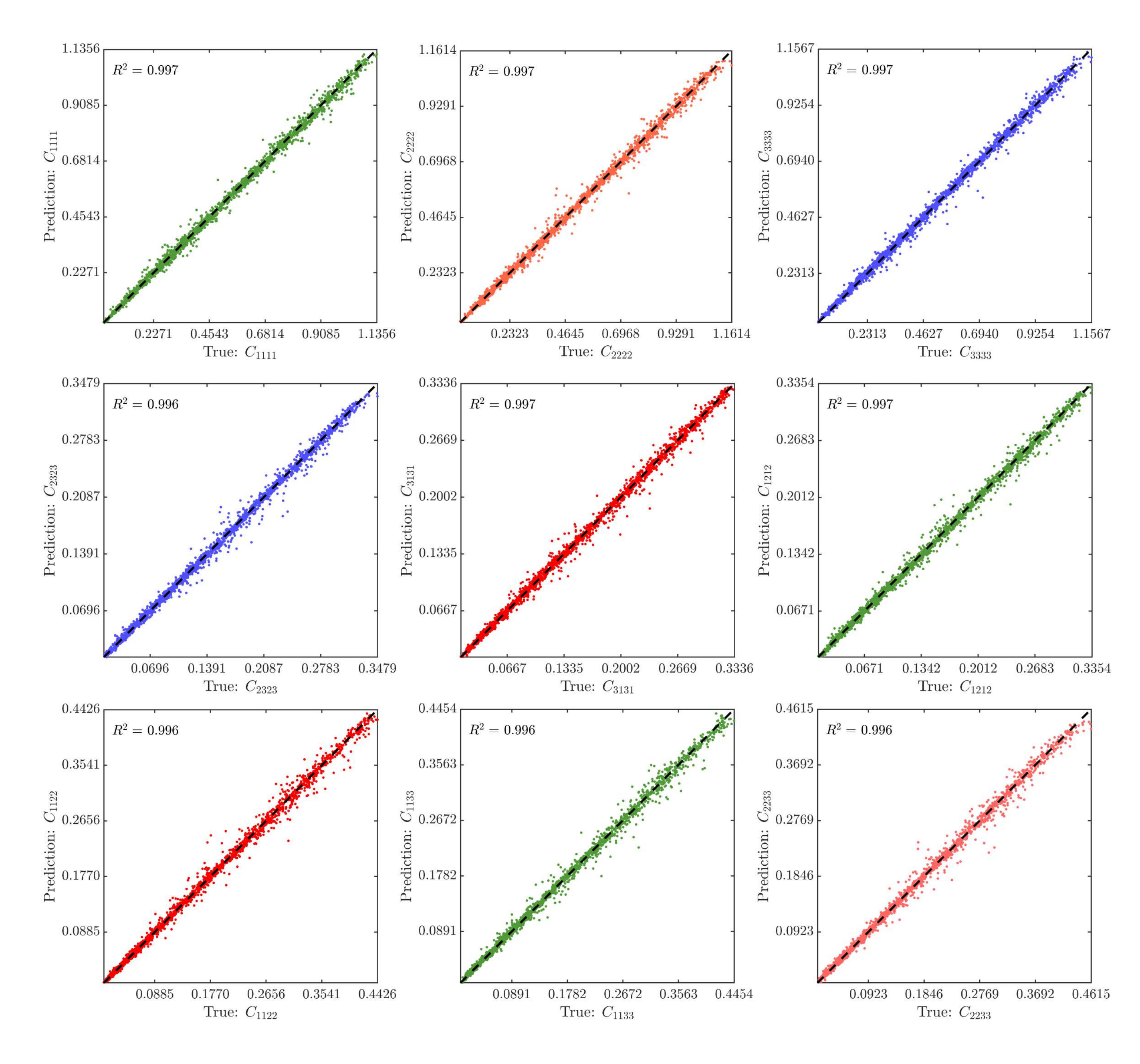}
	\caption{\small DNN-predicted vs.\ true components of $\hat\dsC$ in the test dataset. Dashed lines represent the ideal line with zero-intercept and unit-slope; the corresponding $R^2$-deviations are indicated.
		\label{fig:fNNaccuracy}}
\end{figure}

\section{Results}
\label{sec:Results}

\subsection{Performance validation of the surrogate model}
\label{subsec:nn_validation}

\begin{figure}[tbp]
	\centering
	\includegraphics[width = 0.667\textwidth]{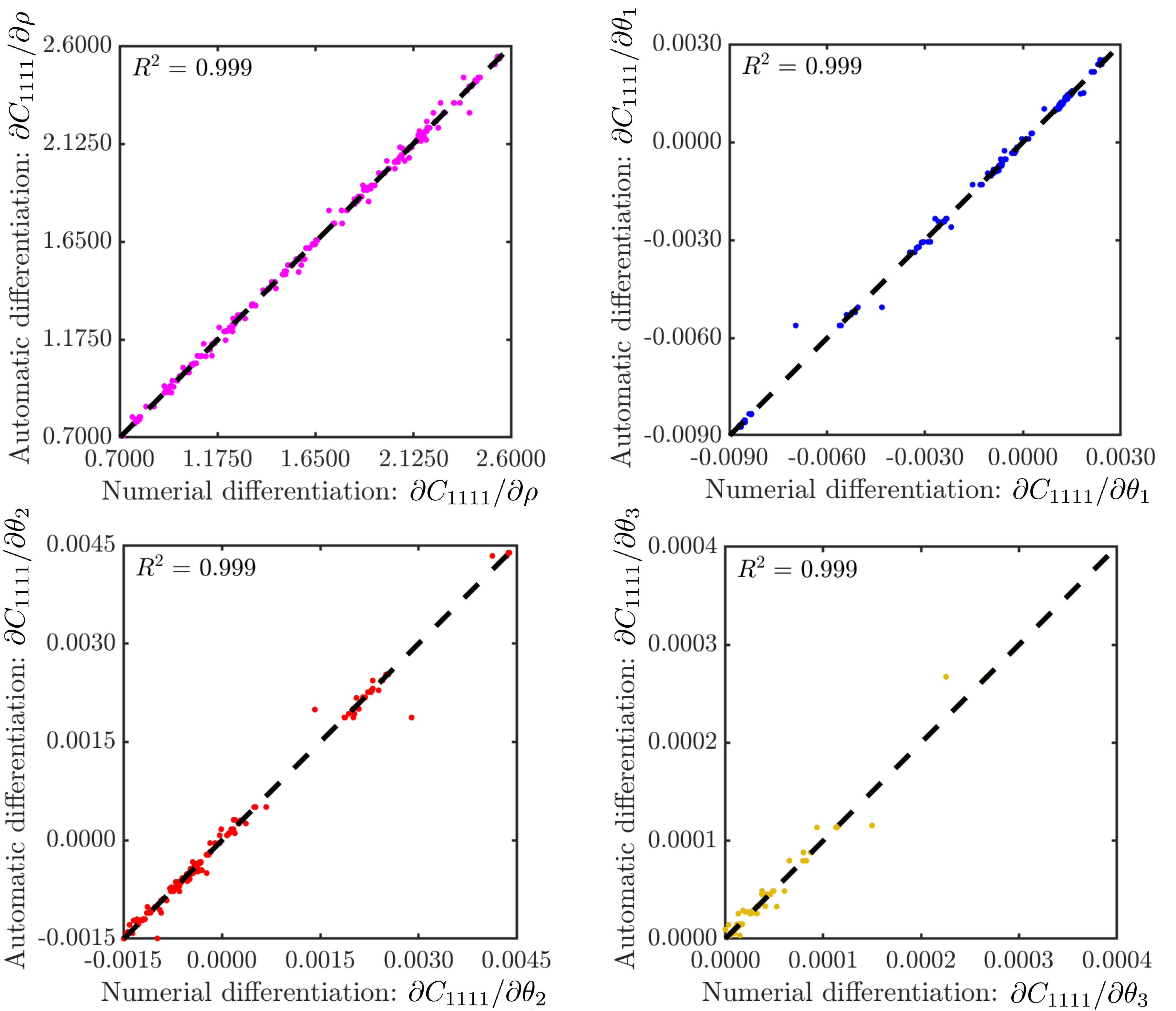}
	\caption{$\mathbb{\hat C}_{1111}$ stiffness sensitivity computed via automatic differentiation vs.\ numerical differentiation in the test set. All dashed lines represent the ideal line with zero-intercept and unit-slope; the corresponding $R^2$-deviations are indicated.}
	\label{fig:sensitivity}
\end{figure}

Before applying the trained surrogate model for the homogenized stiffness to topology optimization problems, we first assess the accuracy of the surrogate model. For a quantitative assessment, the DNN is trained on $n_\text{train}=$19,170 pairs of design parameters and their corresponding effective elastic stiffnesses from FEM, i.e., $(\bfTheta,\bfS)$, followed by testing on an independent test set of containing $n_\text{test}=$2,130 pairs. Ideally, the stiffness prediction $\calF_\omega[\bfTheta]$ from the DNN should agree with the true stiffness, i.e., $\calF_\omega[\bfTheta]\approx \bfS$. Thus, in a plot of predicted vs.\ true stiffness (\figurename\ref{fig:fNNaccuracy}), the predictions are expected to lie on a line with zero-intercept and unit-slope. The trained DNN achieves an $R^2 \geq 0.996$ accuracy for each stiffness component. We also perform a similar assessment  (\figurename \ref{fig:sensitivity}) for the stiffness sensitivities $\calF_\omega'[\bfTheta]$, where we use the central finite difference scheme applied to FEM-computed stiffnesses to validate the derivatives obtained from the DNN via AD. The AD approach shows good agreement with $R^2\geq0.999$ for each stiffness component.

\subsection{Benchmarks}\label{subsec:benchmarks}

We investigate three design examples -- a cantilever beam, an L-shaped structure, and two designs with simultaneous consideration of multiple load cases. As a baseline, for each example we also present the optimal topologies obtained from the classical SIMP method for a homogeneous and isotropic material with Young's modulus $E_s$ and $\nu_s = 0.3$ (which are the same values as those of the base material for spinodoids). The penalty factor for SIMP is chosen as $p=4$. To reduce numerical artifacts due to checkerboard patterns, the filter radius $r_\text{filter}$ in \eqref{eq:filter} is set to $0.075$ for all subsequent simulations \LZ{(thresholding is applied at the end to achieve pure 0-1 solution)}. For simplicity, we avoid units in the following benchmarks by setting $E_s=1$ and normalizing all lengths. To avoid inadmissible design parameters, we choose $\lambda_1 = 600$ and $\lambda_2 = 60 \times (180/\pi)\ \text{rad}^{-1}$ in \eqref{eq:transform} (see \figurename\ref{fig:transform}). In subsequent examples (with the exception of \textit{Benchmark IV}), the small thickness of the chosen design domains (relative to all other dimensions) and the symmetry in boundary conditions results in uniform distributions of the optimal design parameters across the thickness of each design, thus effectively producing 2D topologies on the macroscale with constant thickness.

\subsubsection{Benchmark I: Cantilever beam}

\begin{figure}[tbp]
	\centering
	\begin{subfigure}{0.45\textwidth}
		\includegraphics[width = \textwidth]{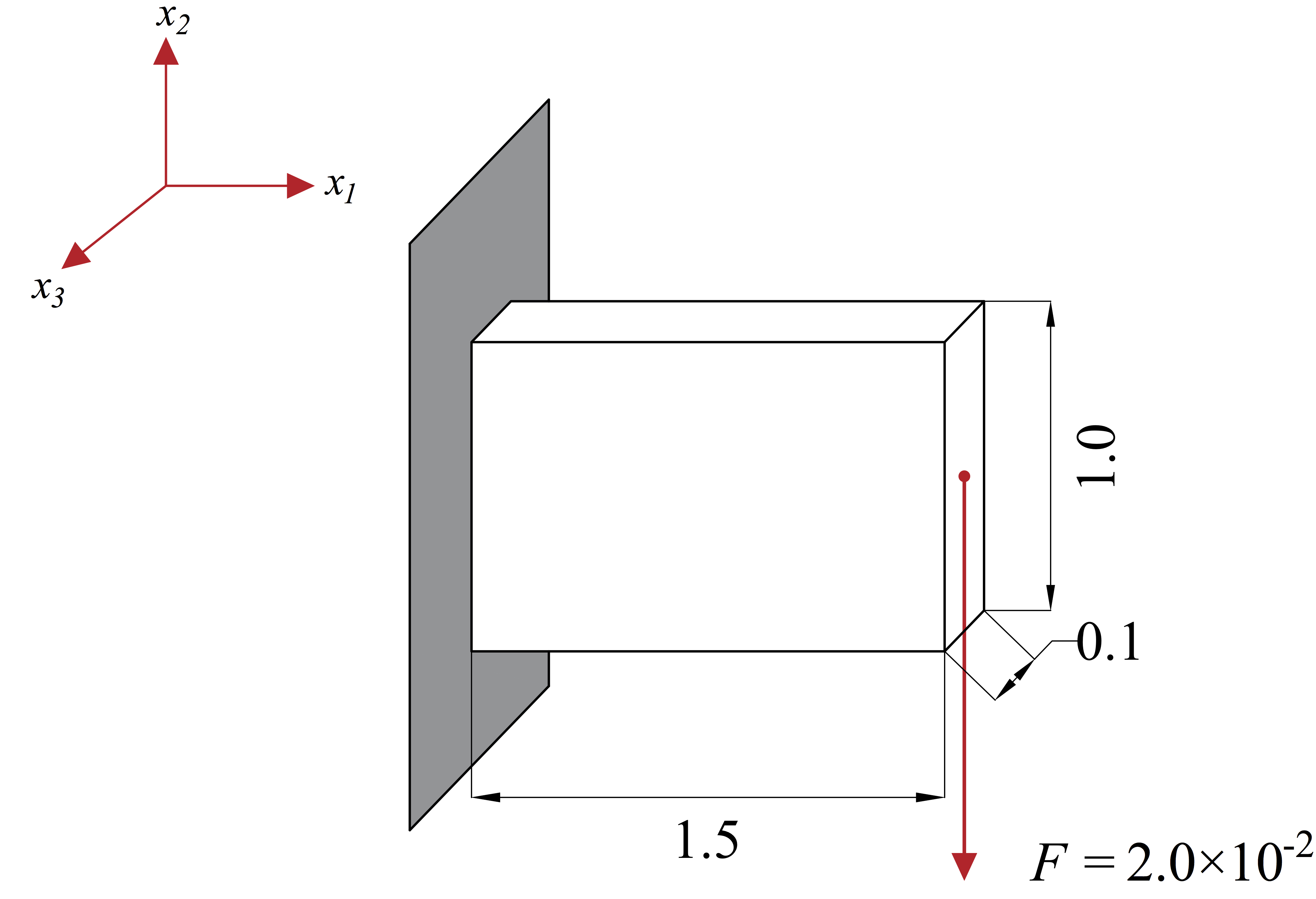}
		\caption{}\label{fig:cantilever}
	\end{subfigure}
\hfill
	\begin{subfigure}{0.5\textwidth}
		\centering
		\includegraphics[width = 0.7\textwidth]{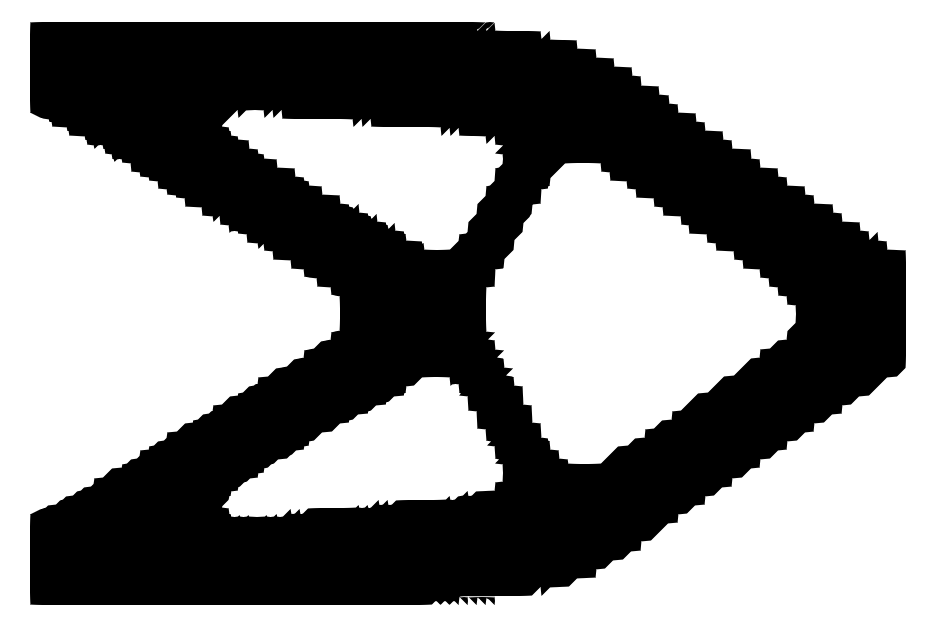}
		\caption{}\label{fig:cantiSIMP}
	\end{subfigure}

	\begin{subfigure}{0.59\textwidth}
		\centering
		\begin{tabular}{cccccccc}
			\toprule
			& \textbf{Spinodoid} & \textbf{SIMP} \\
			\midrule
			Optimized compliance &  $5.602 \times 10^{-1}$ & $\LZ{5.299 \times 10^{-1}}$\\
			\toprule
		\end{tabular}
		\caption{}
		\label{tab:cantiObj}
	\end{subfigure}
	\caption{\textit{Benchmark I:} (a) Schematic of the end-loaded cantilever beam whose compliance is to be optimized. \LZ{(b) Front view of the optimal topology obtained using the SIMP method.} (c) Comparison of the optimal compliance for the spinodoid architecture on the microscale (via proposed method) and a solid material (via SIMP).}
\end{figure}

We consider a cantilever beam (\figurename \ref{fig:cantilever}) of size $1.5\times 1.0\times 0.1$ discretized by a mesh of linear tetrahedral elements on a $64\times 48 \times 2$ uniform grid. The left vertical face of the beam is clamped (i.e., displacements are suppressed in all directions). A single point load of magnitude $2.0 \times 10^{-2}$ in the downward direction is applied at the center of the right vertical face. We seek to minimize the compliance subject to an average relative density  constraint of $\bar \rho = 0.5$. As an initial guess, all elements are initialized with spinodoid relative density $\rho=0.5$, anisotropy parameters  $(\theta_1, \theta_2, \theta_3) = (15^\circ, 0^\circ, 0^\circ)$ and orientation $\alpha=0^\circ$. The initial design is intentionally chosen to be  lamellar-type (with all lamellae normal to length of the beam). Such a design will strongly deform under axial loads (as evident from their low Young's moduli; see {\figurename \ref{subfig:lamellar}}) and therefore show a highly compliant behavior.

\begin{figure}
	\centering
	\begin{subfigure}{\textwidth}
		\centering
		\includegraphics[width = \linewidth]{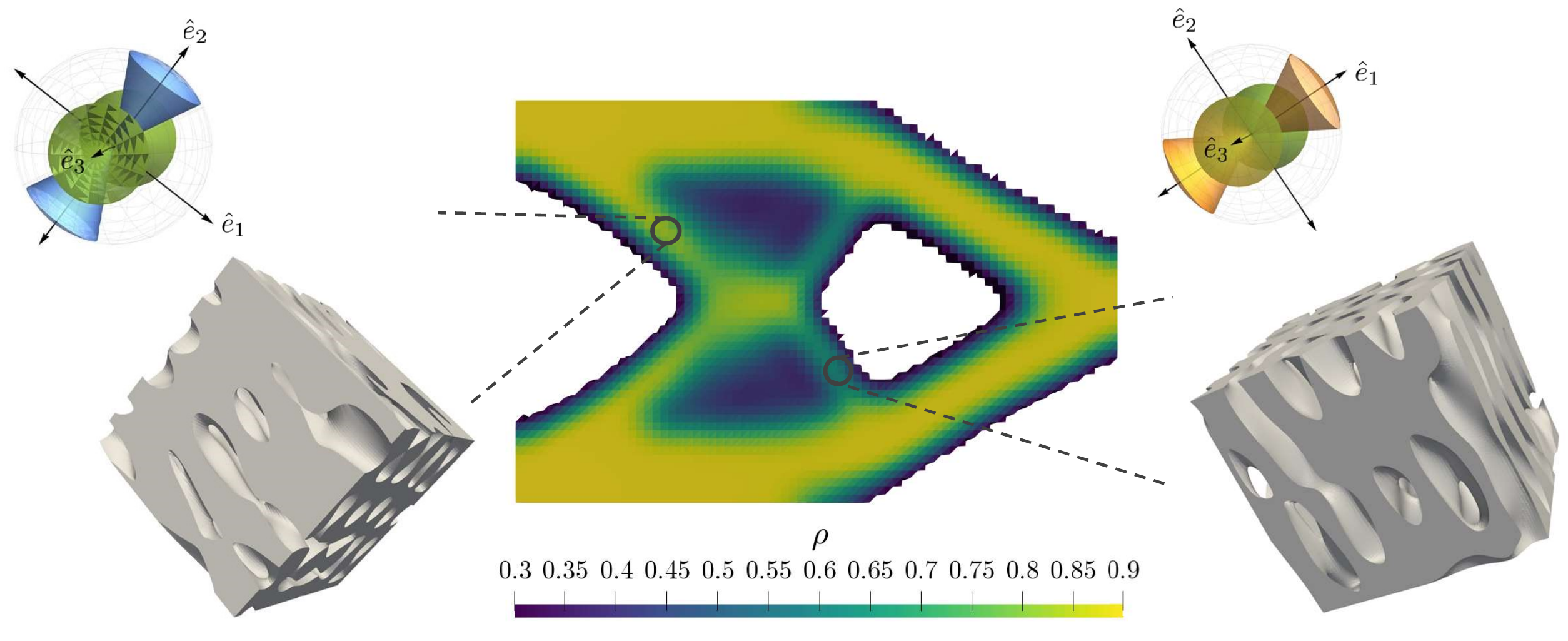}
		\caption{Optimized relative density $\rho$}\label{fig:cantiRho}
	\end{subfigure}

	\begin{subfigure}{0.4\textwidth}
		\centering
		\includegraphics[width = \linewidth]{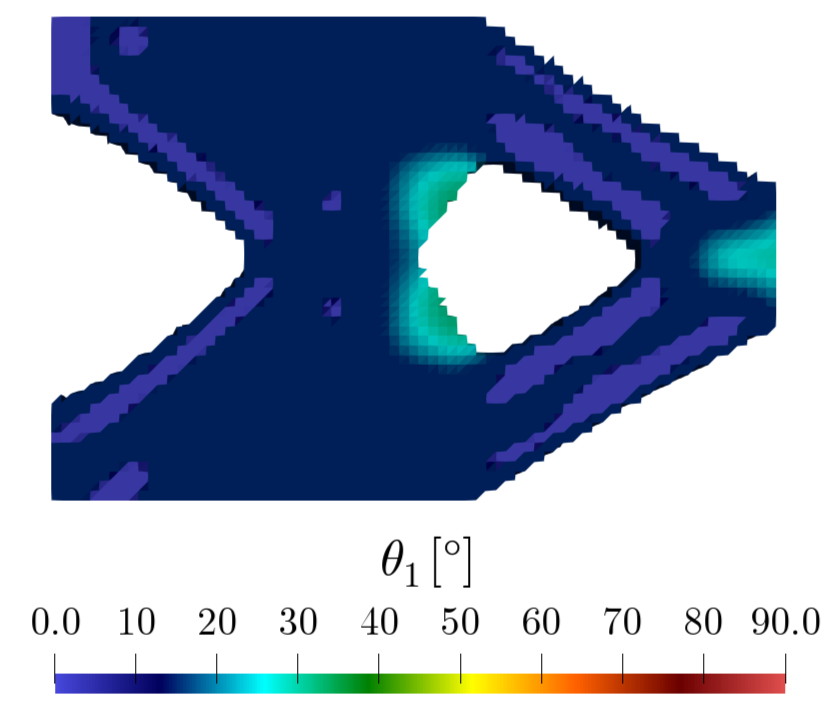}
		\caption{Optimized spinodoid parameter $\theta_1$}
	\end{subfigure}
	\qquad
	\begin{subfigure}{0.4\textwidth}
		\centering
		\includegraphics[width = \linewidth]{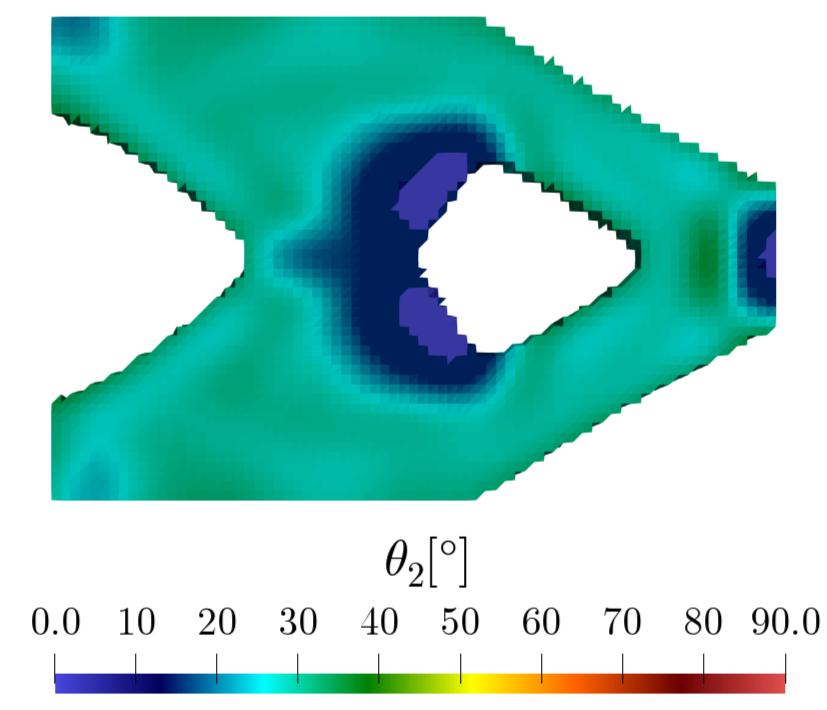}
		\caption{Optimized spinodoid parameter $\theta_2$}
	\end{subfigure}

	\begin{subfigure}{0.4\textwidth}
		\centering
		\includegraphics[width = \linewidth]{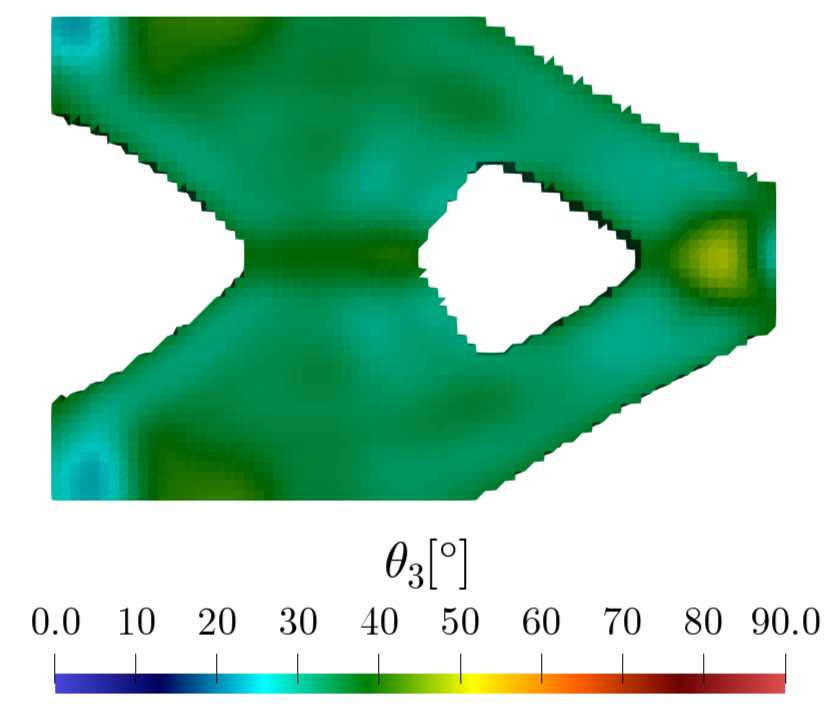}
		\caption{Optimized spinodoid parameter $\theta_3$}
	\end{subfigure}
	\qquad
	\begin{subfigure}{0.4\textwidth}
		\centering
		\includegraphics[width = \linewidth]{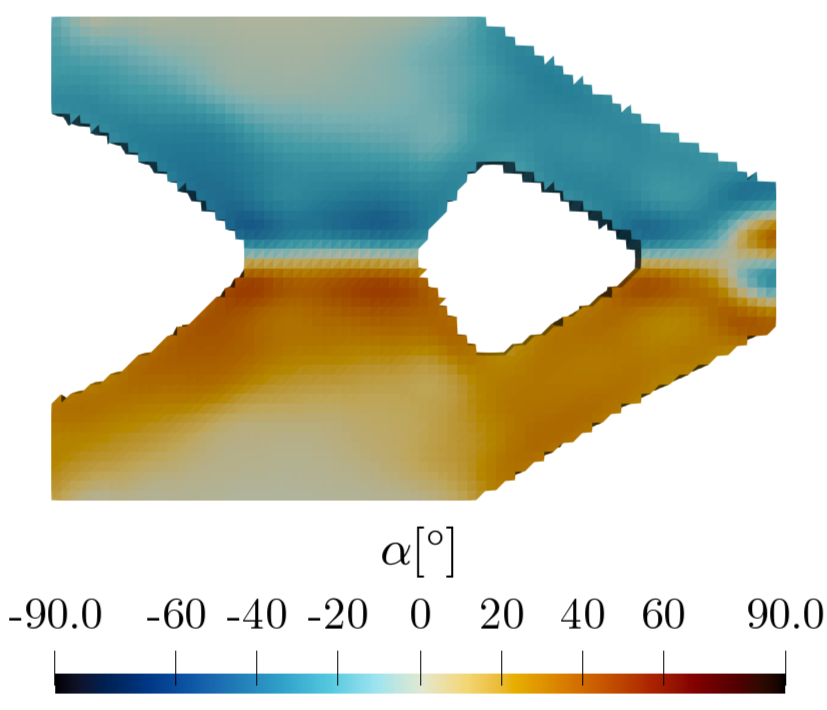}
		\caption{Optimized material orientation $\alpha$}
		\label{fig:cantiAlpha}
	\end{subfigure}
	\caption{\textit{Benchmark I:} Front view of the optimal topology (material distribution, anisotropy, and orientation) of the cantilever beam with spinodoid microscale architectures. The spatially-varying design parameters are illustrated across the body, and two examples of distinct microscale topologies are shown in (a).}
	\label{fig:cantiResults}
\end{figure}

\begin{figure}
	\centering
	\includegraphics[width = \textwidth]{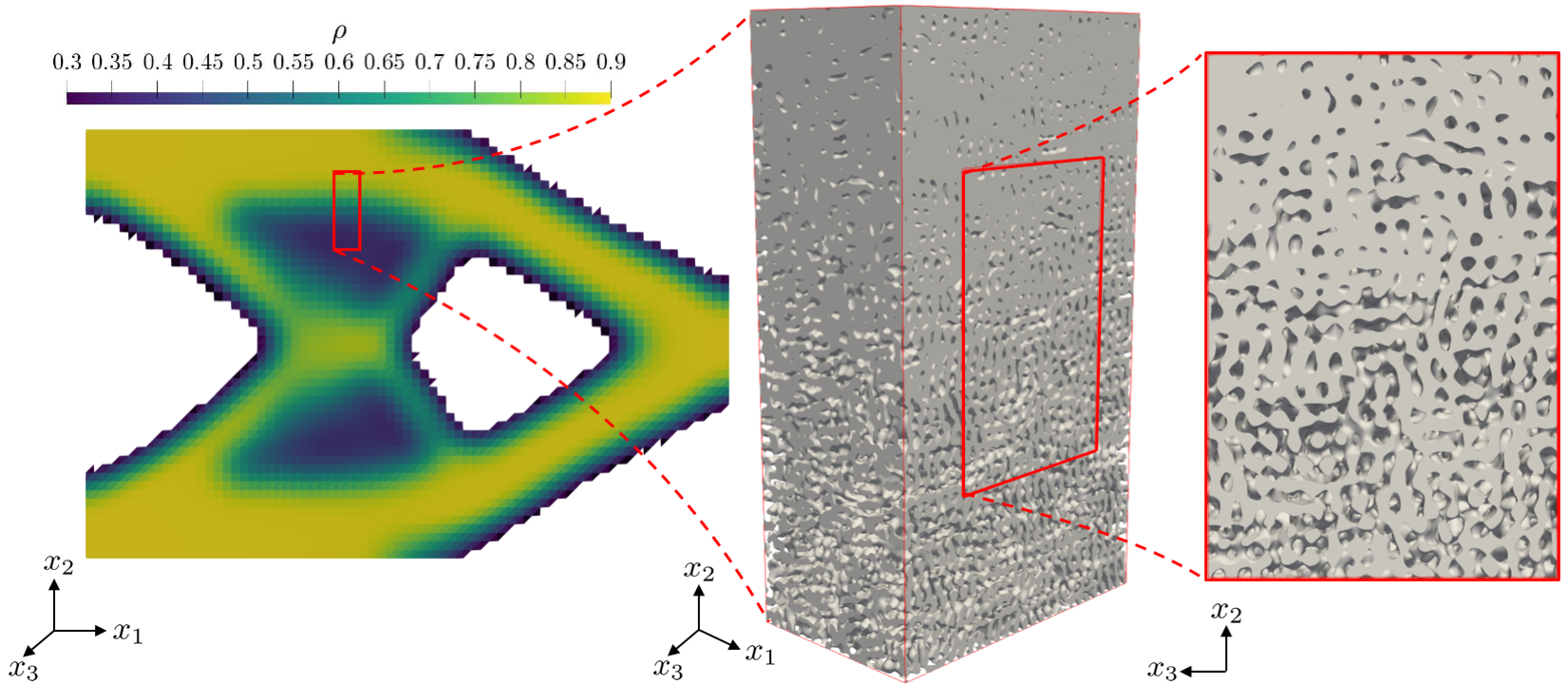}
	\caption{\textit{Benchmark I:} Seamlessly spatially-variant spinodoid topology with fully resolved microstructure in a subdomain of the optimized cantilever beam design. The microscale GRF is generated with wavenumber $\beta=1600\pi/3$ (arbitrarily chosen and sufficiently large under the assumption of an approximate separation of scales). Details on generating the microscale topology are presented in \ref{sec:spatiallyVariantAppendix}.} \label{fig:cantiSpatiallyVariant}
\end{figure}

The optimal design with anisotropic spinodoid architectures is shown in \figurename\ref{fig:cantiResults}. The material distribution (\figurename \ref{fig:cantiRho}) resembles the optimal topology obtained via SIMP (\figurename \ref{fig:cantiSIMP}), albeit the former is characterized by larger regions with intermediate density. \LZ{The optimal compliances obtained by the single-scale SIMP design and our multiscale spinodoid design are shown in \figurename\ref{tab:cantiObj}.} Contrary to the initial guess of a lamellar topology, the final design is dominated by cubic topologies with larger  Young's moduli along the principal directions. Throughout the macroscale body, the spinodoid microstructures are rotated, so that their preferred orientations follow the material distribution (see \figurename\ref{fig:cantiRho} and \figurename\ref{fig:cantiAlpha}). \mbox{\figurename\ref{fig:cantiSpatiallyVariant}} illustrates the seamlessly spatially-variant spinodoid topology (with fully resolved microstructure), which bypasses the challenge of incompatible microstructures in periodic metamaterials. To avoid the high computational expense of generating a large mesh with fully resolved microstructure, the microscale topology is illustrated for a representative subdomain only.

\subsubsection{Benchmark II: L-shaped structure}

\begin{figure}[t]
	\centering
	\begin{subfigure}{0.5\textwidth}
		\centering
		\includegraphics[width = \textwidth]{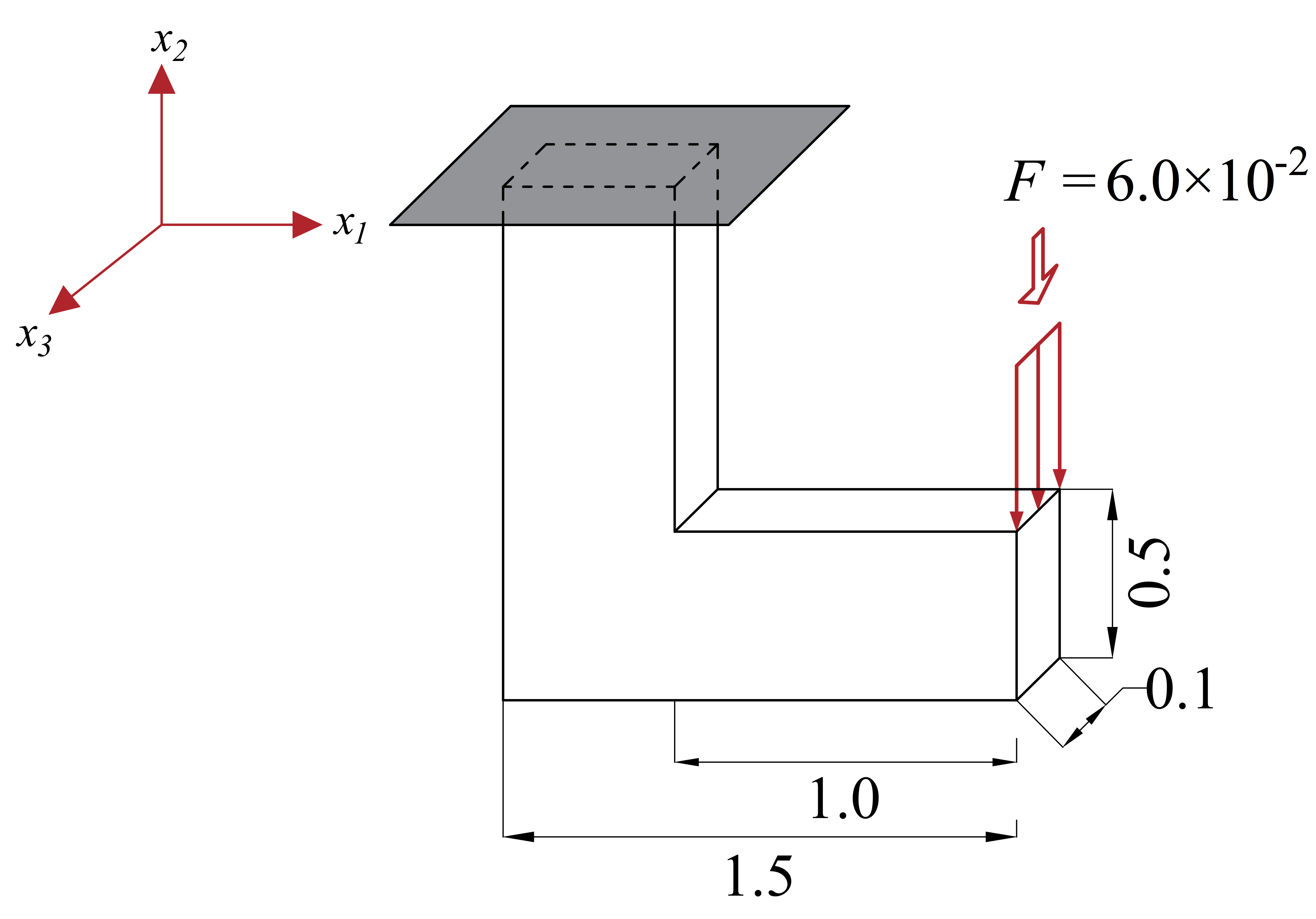}
		\caption{}\label{fig:Lshape}
	\end{subfigure}
	\hfill
	\begin{subfigure}{0.32\textwidth}
		\centering
		\includegraphics[width =\textwidth]{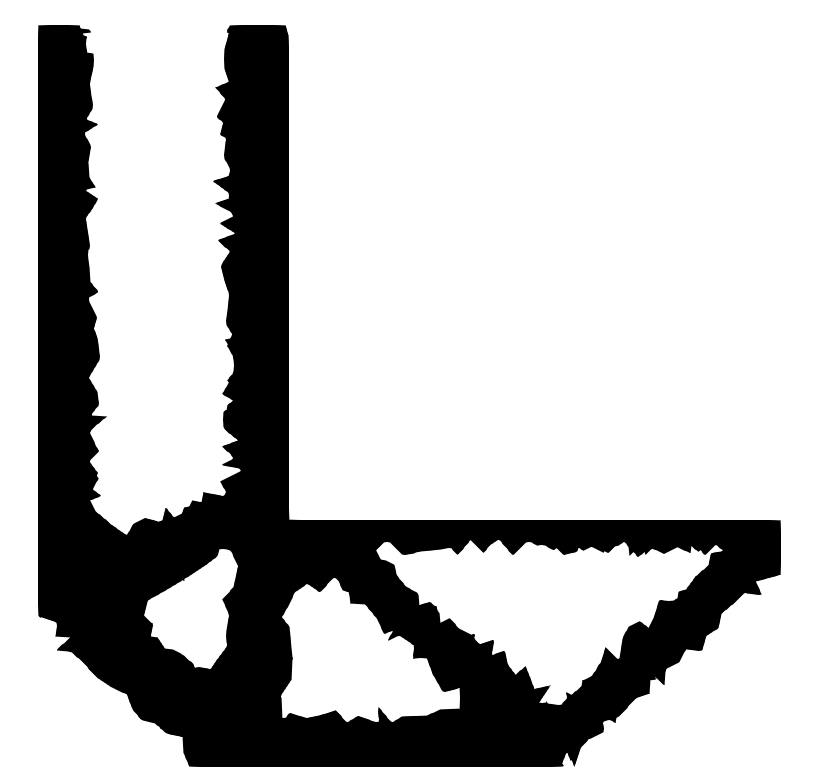}
		\caption{}
		\label{fig:LshapeSIMP}
	\end{subfigure}

	\centering
	\begin{subfigure}{\textwidth}
		\begin{tabular}{cccccclc}
			\toprule
			& \multicolumn{5}{c}{\textbf{Spinodoid}} &  & \textbf{SIMP} \\ \cmidrule{1-6} \cmidrule{8-8}
			No. of elements & 6870 & 8090 & 14610 & 22060 & 26341 &  & 22060 \\
			Optimized compliance & $2.429\times 10^1$ & $2.459\times 10^1$ & $2.425\times 10^1$ & \bm{$2.282\times 10^1$} & $2.308\times 10^1$ &  & \LZ{$2.490\times 10^1$} \\
			\toprule
		\end{tabular}
		\caption{}\label{tab:LshapeObj}
	\end{subfigure}
	\caption {\textit{Benchmark II:} (a) Schematic of the L-shaped structure whose compliance is to be optimized. \LZ{(b) Front view of the optimal topology obtained using the SIMP method.} (c) Comparison of the optimal compliance with spinodoid microscale architecture (via the proposed method for different mesh resolutions) and solid material (via SIMP).}
\end{figure}

In this example, an L-shape structure with the dimensions shown in \figurename\ref{fig:Lshape} is optimized. The top face is fixed in all directions, while a uniformly-distributed vertical load is applied on the lower right edge, as shown. We seek to minimize the total compliance subject to an average relative density constraint of $\bar \rho = 0.5$. As an initial guess, all elements are initialized with a spinodoid relative density $\rho=0.5$, anisotropy parameters  $(\theta_1, \theta_2, \theta_3) = (35^\circ, 15^\circ, 15^\circ)$ (cubic topology), and orientation $\alpha=0^\circ$. The domain is discretized into a finite element mesh with 22,060 linear tetrahedral elements, yielding a total of 110,300 design variables.

\begin{figure}
	\centering
	\begin{subfigure}{0.9\textwidth}
		\centering
		\includegraphics[width = \linewidth]{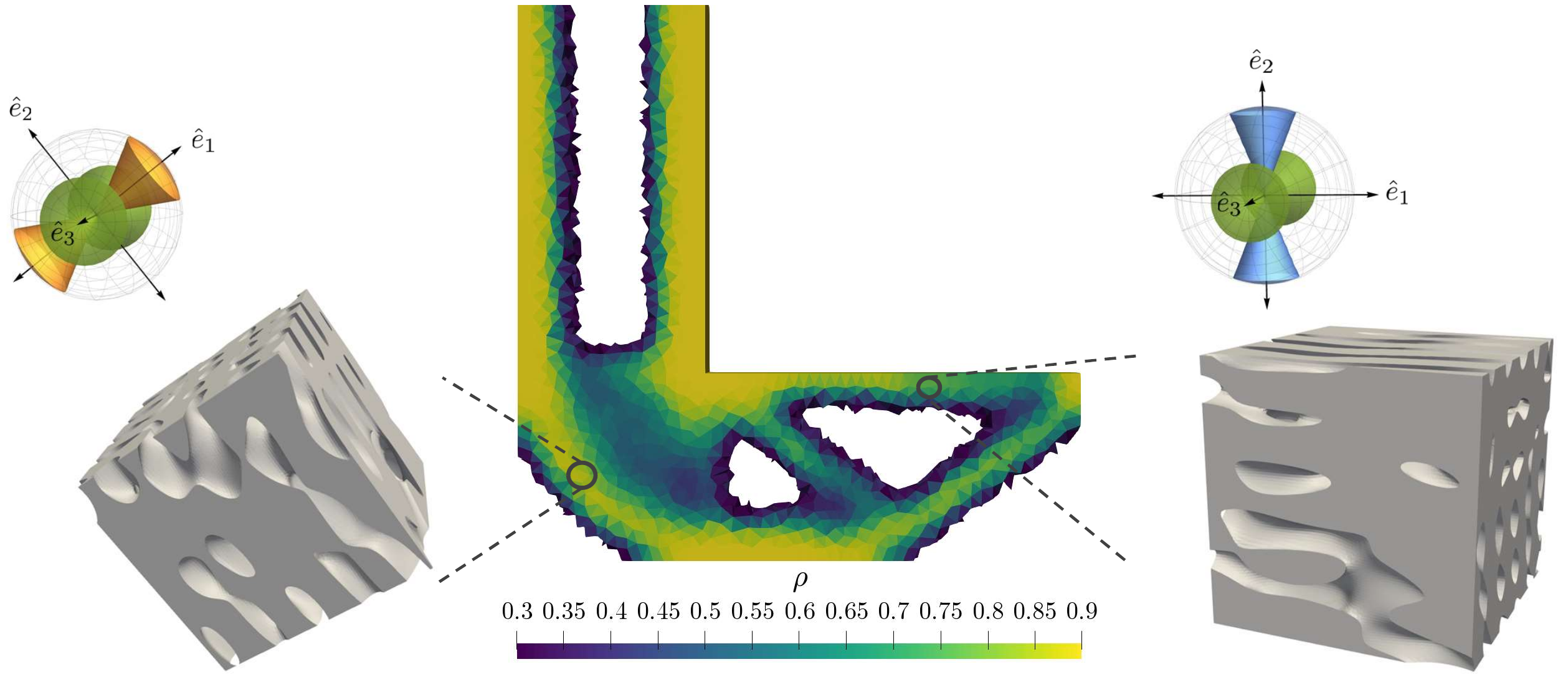}
		\caption{Optimized relative density: $\rho$} \label{fig:LshapeRho}
	\end{subfigure}
	\begin{subfigure}{0.32\textwidth}
		\centering
		\includegraphics[width = \linewidth]{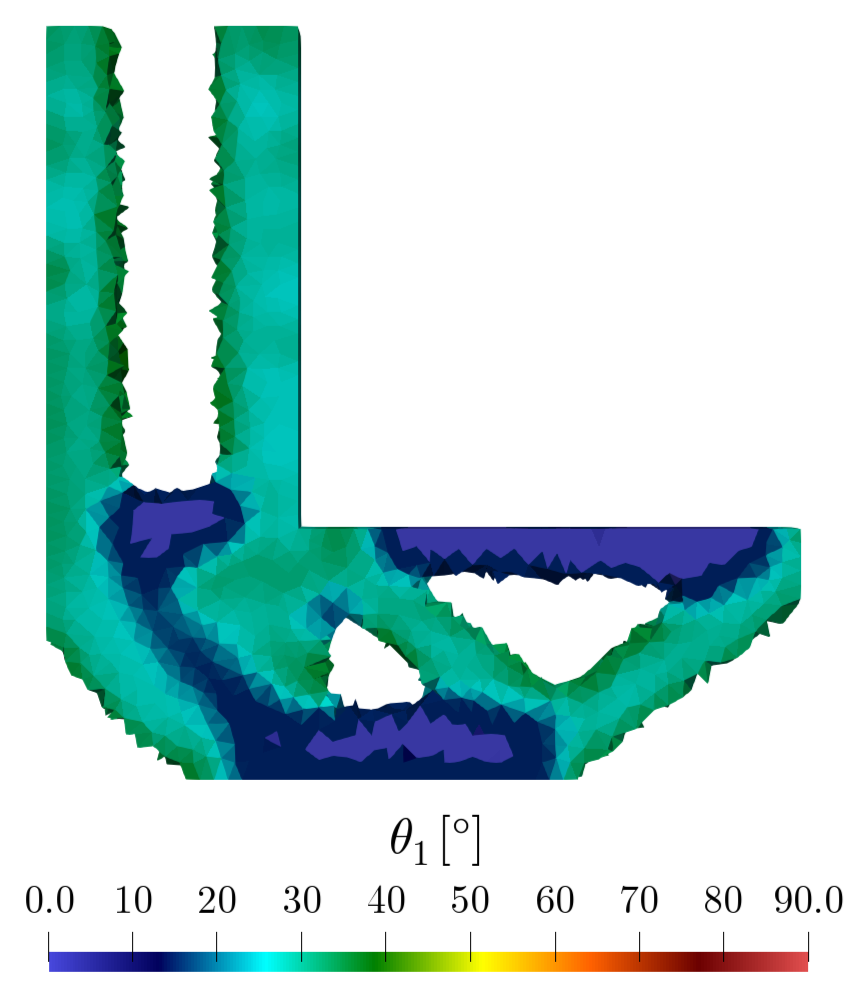}
		\caption{Optimized spinodoid parameter: $\theta_1$}
	\end{subfigure}
	\qquad \qquad
	\begin{subfigure}{0.32\textwidth}
		\centering
		\includegraphics[width = \linewidth]{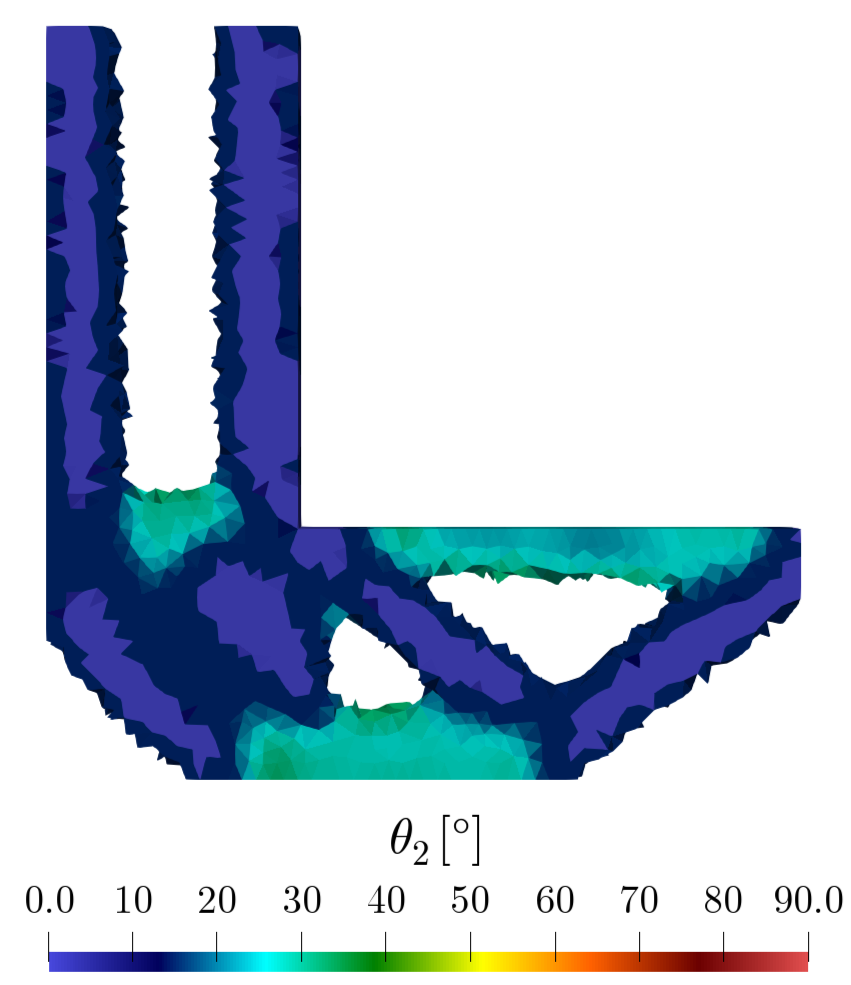}
		\caption{Optimized spinodoid parameter: $\theta_2$}
	\end{subfigure}

	\begin{subfigure}{0.32\textwidth}
		\centering
		\includegraphics[width = \linewidth]{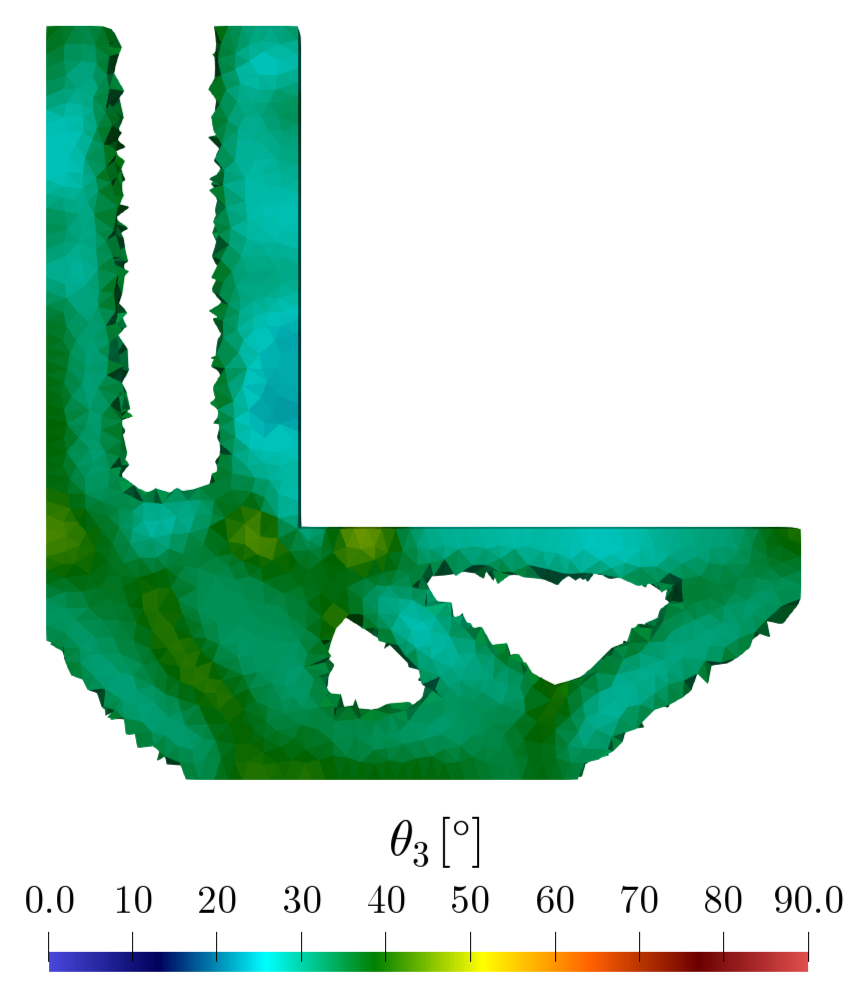}
		\caption{Optimized spinodoid parameter: $\theta_3$}
	\end{subfigure}
	\qquad \qquad
	\begin{subfigure}{0.32\textwidth}
		\centering
		\includegraphics[width = \linewidth]{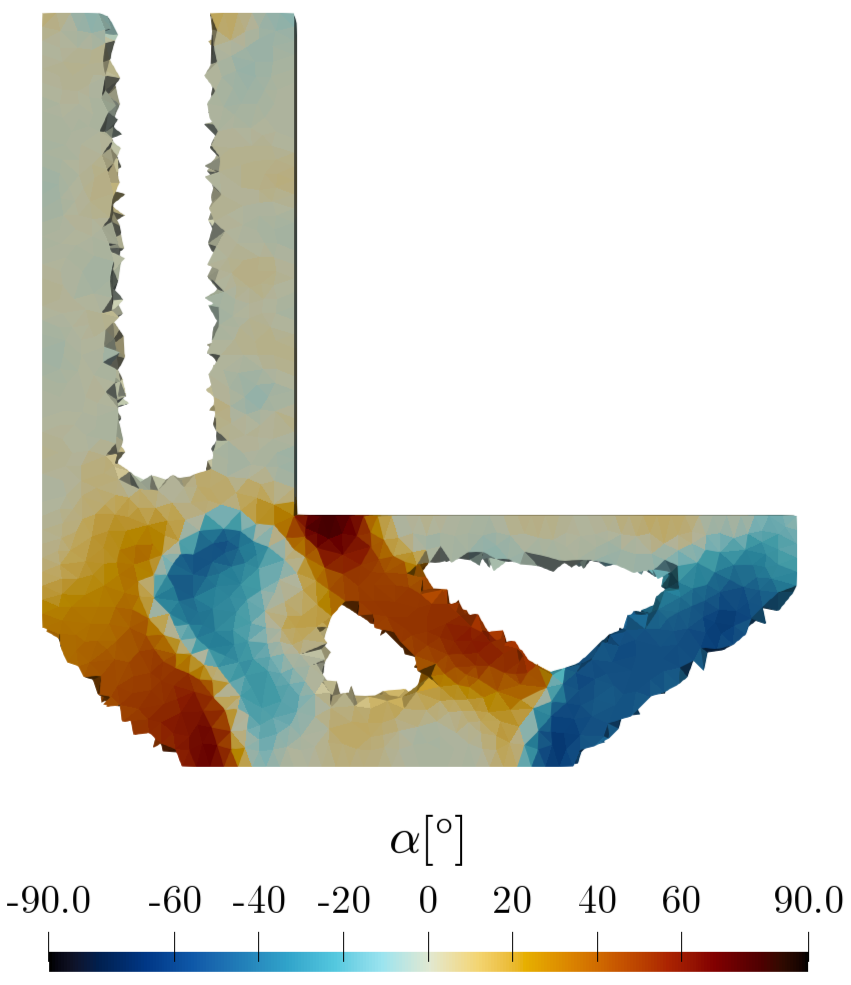}
		\caption{Optimized material orientation: $\alpha$}
		\label{fig:LshapeAlpha}
	\end{subfigure}
	\caption{\textit{Benchmark II:} Front view of the optimal topology (material distribution, anisotropy, and orientation) of the L-shaped structure with spinodoid microscale architecture. Examples of the spatially-varying design parameters and the corresponding microscale topologies are shown in~(a).}
	\label{fig:LshapeResults}
\end{figure}

The optimal topology with anisotropic spinodoid architectures is shown in \figurename\ref{fig:LshapeResults} and resembles the SIMP result (\figurename \ref{fig:LshapeSIMP}) in terms of the material distribution. The spinodoid-based design achieves \LZ{$8.35\%$} (\figurename \ref{tab:LshapeObj}) improvement in the compliance relative to the SIMP-based design. This is possible due to the spatially-varying anisotropy; e.g., the zoomed-in microstructures in \figurename\ref{fig:LshapeRho} show columnar features, which provide high stiffness in the principal stress directions -- thus orienting material not only at the macroscale but also on the microscale for optimal compliance. \figurename\ref{tab:LshapeObj} lists the optimal compliance for five different mesh resolutions. Increasing the number of finite elements does not affect the optimized compliance significantly. Counter-intuitively, higher mesh resolution may not necessarily yield a lower compliance in the context of a multiply-connected design parameter space and non-convex property space -- e.g.\ for spinodoids, $\rho\in\{0\}\cup[\rho_{\text{min}},1]$ and $\theta_1,\theta_2,\theta_3\in\{0\}\cup[0,\pi/2]$. Similar observations were reported and explained previously by, e.g., \citet{BaiZhu2016} and \citet{KumarNonConvex}.

\subsection{Benchmark III: Multiple load cases -- symmetric loading}
\label{sec:Benachmark3}

Here, we extend our multiscale topology optimization algorithm for compliance minimization to the simultaneous consideration of multiple load cases. The objective function in \eqref{eq:compliance} is modified as the sum of the compliances for each individual load case, i.e.,
\begin{equation}\label{eq:objMulti}
\Phi(x) = \sum^{M}_{i=1}\boldface{U}^{i}(x)^T\boldface{K}^i(x)\boldface{U}^i(x),
\end{equation}
where the superscript $(\cdot)^i$ denotes the $i^\text{th}$ load case, and $M$ is the total number of load cases considered.

\begin{figure}[t]
	\centering
	\begin{subfigure}{0.45\textwidth}
		\centering
		\includegraphics[width = \textwidth]{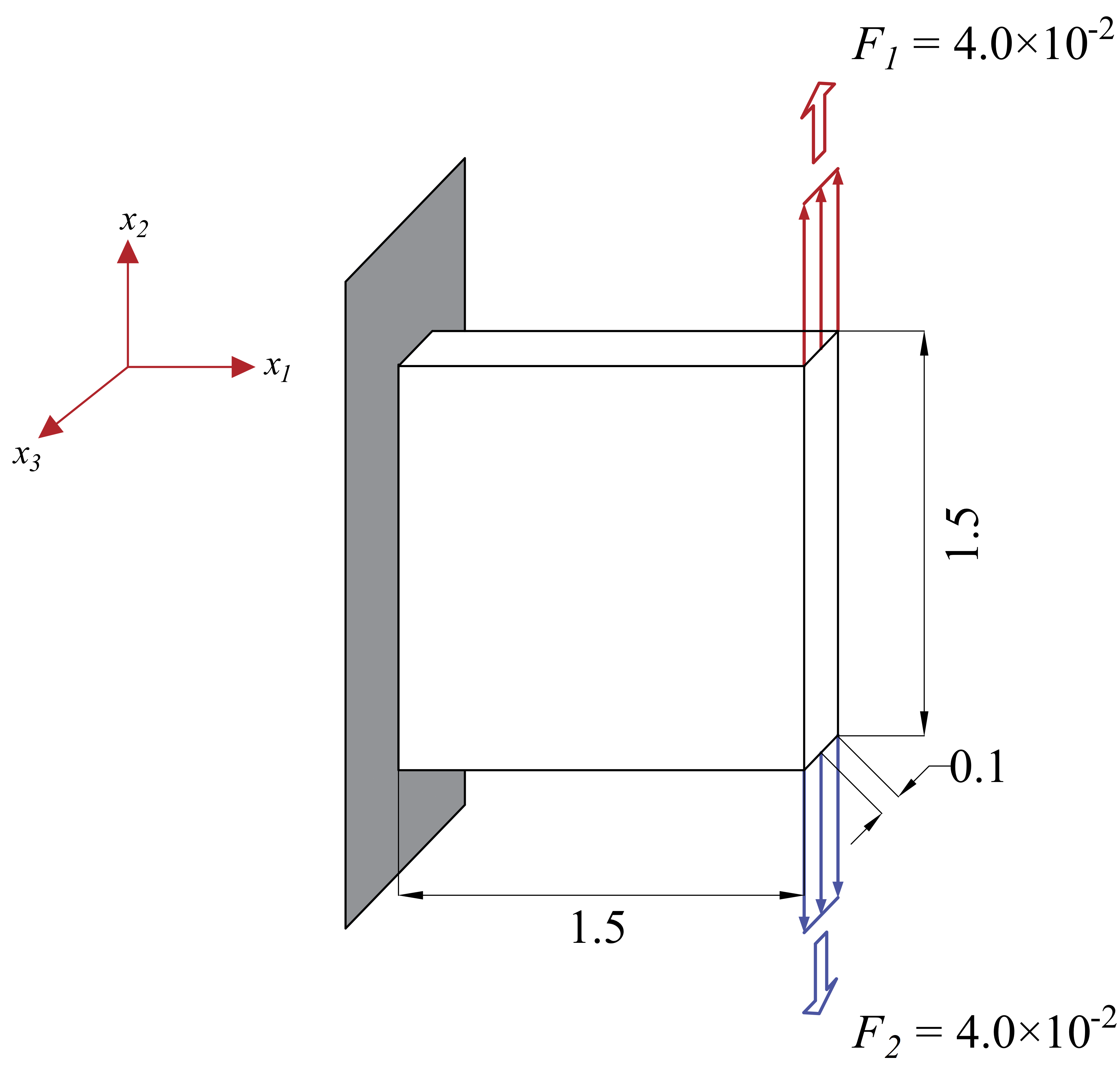}
		\caption{}\label{fig:multiSymm}
	\end{subfigure}
	\qquad \qquad
	\begin{subfigure}{0.35\textwidth}
		\centering
		\includegraphics[width =\textwidth]{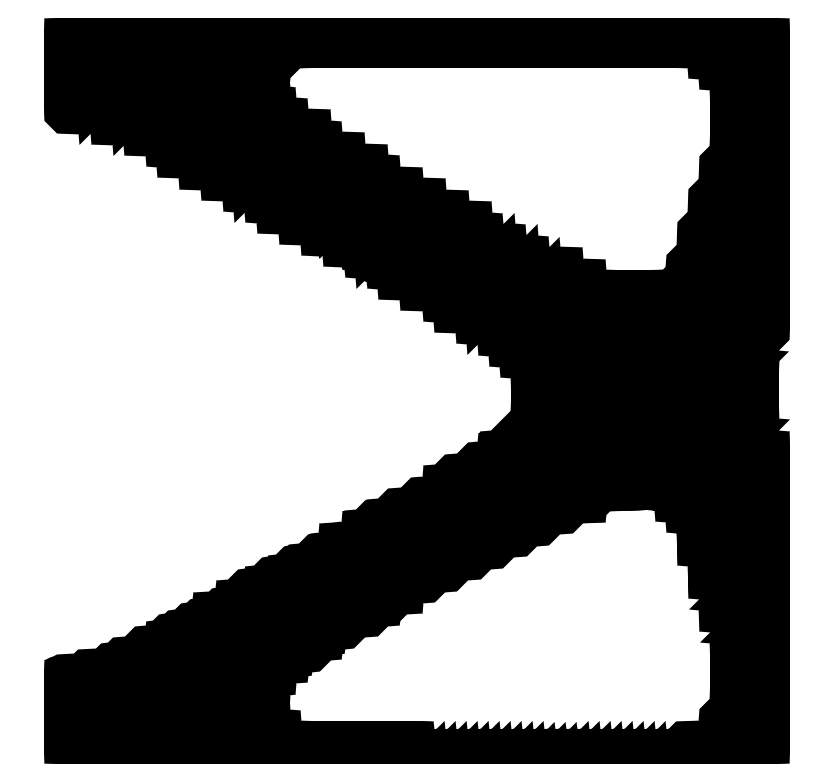}
		\caption{}
		\label{fig:multiSymmSIMP}
	\end{subfigure}

	\centering
	\begin{subfigure}{\textwidth}
		\begin{tabular}{cccccclc}
			\toprule
			& \multicolumn{5}{c}{\textbf{Spinodoid}} &  & \textbf{SIMP} \\ \cmidrule{1-6} \cmidrule{8-8}
			No. of elements & 12150 & 16224 & 21600 & 25350 & 31104 &  & 25350 \\
			Optimized compliance & $8.101 \times 10^{1}$ & $8.412 \times 10^{1}$ & $8.340 \times 10^{1}
			$ & \bm{$8.297 \times 10^{1}}
			$ & $8.282 \times 10^{1}$ &  & \LZ{$8.723 \times 10^{1}$} \\
			\toprule
		\end{tabular}
		\caption{}\label{tab:multiSymmObj}
	\end{subfigure}
	\caption {\textit{Benchmark III:} (a) Schematic of the structure to be optimized for two symmetric load cases (shown by the two sets of applied forces). \LZ{(b) Front view of the optimal topology obtained using the SIMP method.} (c) Comparison of the optimal compliance with spinodoid microscale architectures (via the proposed method for different mesh resolutions) and solid material (via SIMP).}
\end{figure}

As an example, we consider the beam shown in \figurename\ref{fig:multiSymm} with $M=2$ load cases: a uniformly-distributed force in the upward direction on the upper-right edge, and a uniformly-distributed force in the downward direction on the lower-right edge. The average relative density constraint remains $\bar\rho=0.5$. As an initial guess, all elements are initialized with a spinodoid relative density $\rho=0.5$, anisotropy parameters $(\theta_1, \theta_2, \theta_3) = (35^\circ, 15^\circ, 15^\circ)$ (cubic topology), and orientation $\alpha=0^\circ$. The domain is discretized into a uniform linear tetrahedral mesh with 25,350 elements and 126,750 design variables  \revised{(a convergence study (\figurename\ref{tab:multiSymmObj}) revealed that increasing the mesh resolution leads to only marginal changes in accuracy; e.g., increasing the number of elements 23\% (from 25,350 to 31,104) leads to a change in the optimal compliance of approximately 0.2\%)}.

\begin{figure}
	\centering
	\begin{subfigure}{\textwidth}
		\centering
		\includegraphics[width = \textwidth]{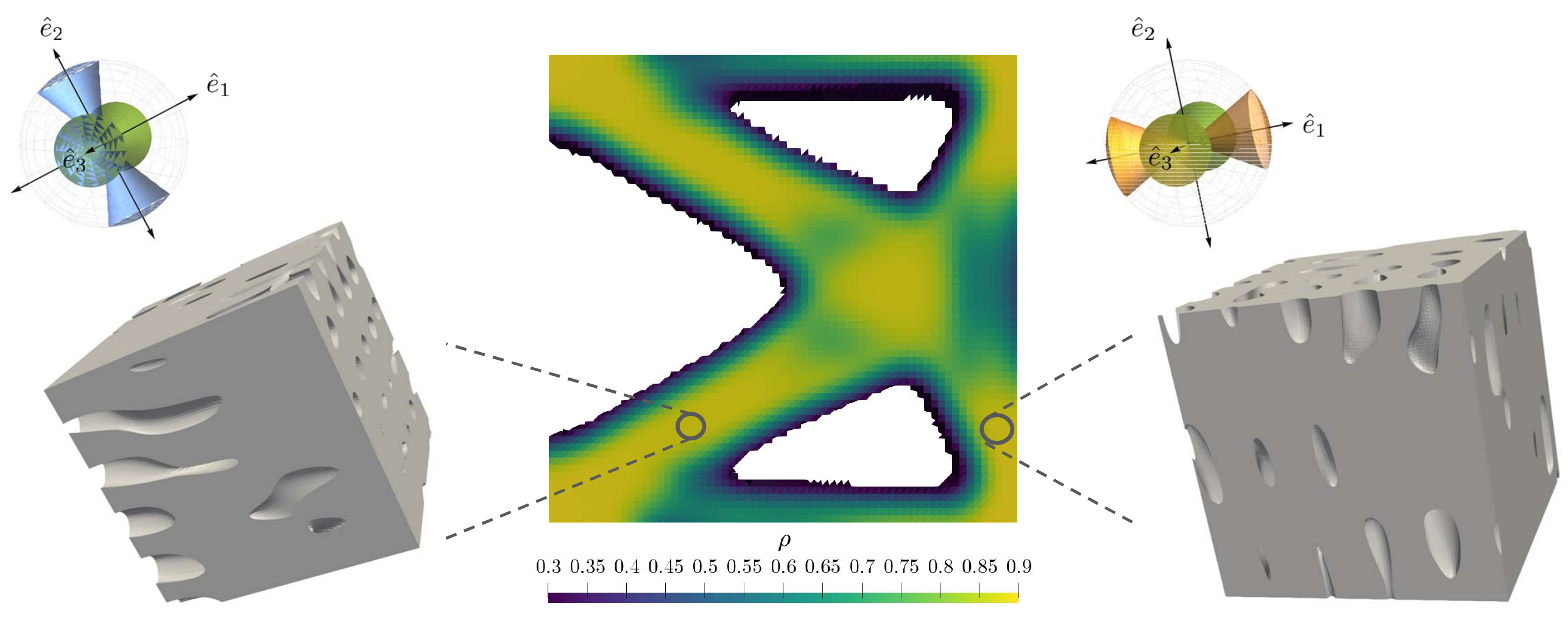}
		\caption{Optimized relative density $\rho$} \label{fig:multiSymmRho}
	\end{subfigure}
	\begin{subfigure}{0.32\textwidth}
		\centering
		\includegraphics[width = \linewidth]{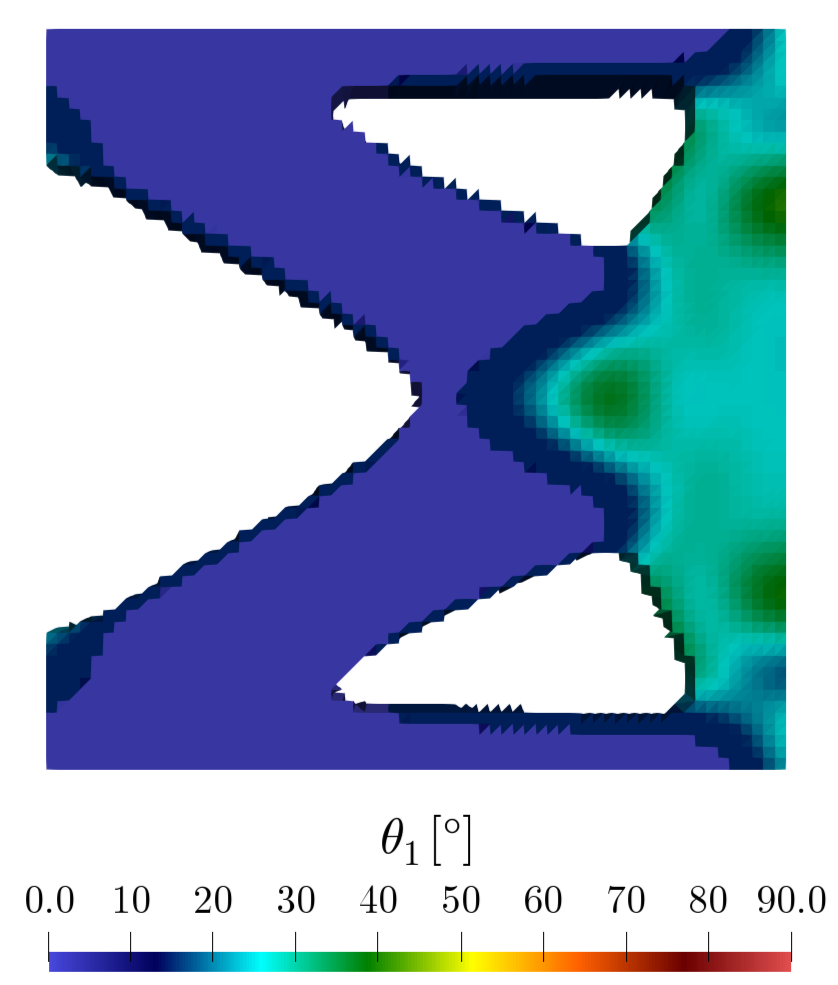}
		\caption{Optimized spinodoid parameter $\theta_1$}
	\end{subfigure}
	\qquad \qquad
	\begin{subfigure}{0.32\textwidth}
		\centering
		\includegraphics[width = \linewidth]{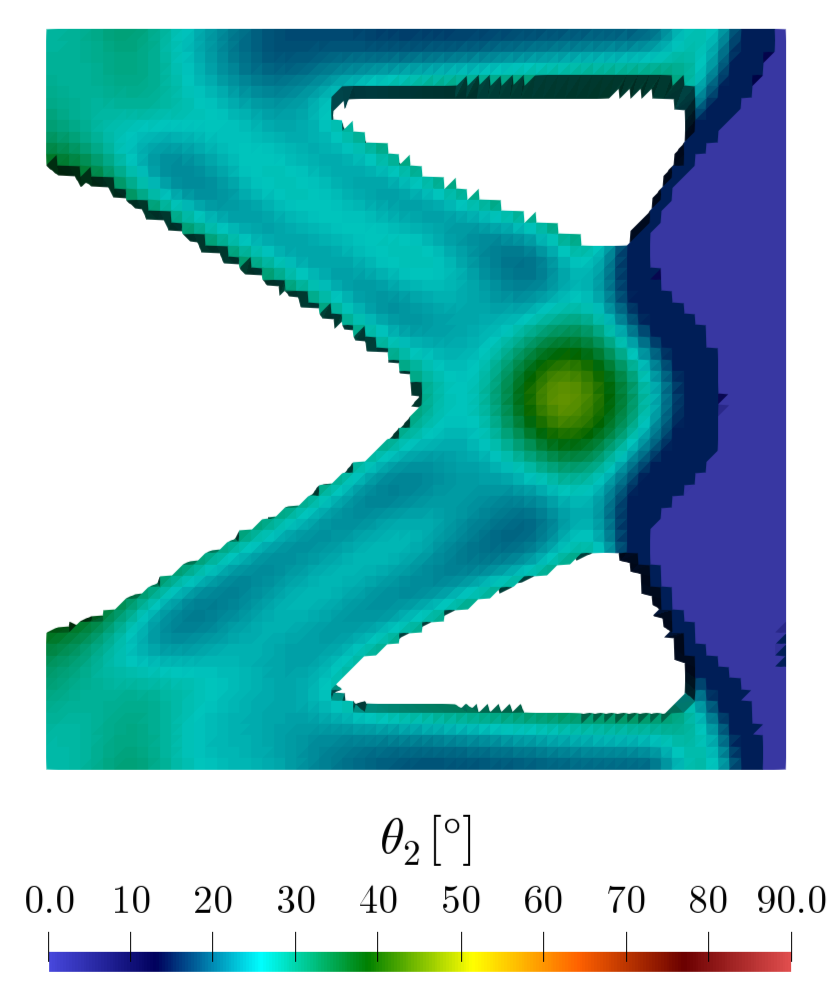}
		\caption{Optimized spinodoid parameter $\theta_2$}
	\end{subfigure}

	\begin{subfigure}{0.32\textwidth}
		\centering
		\includegraphics[width = \linewidth]{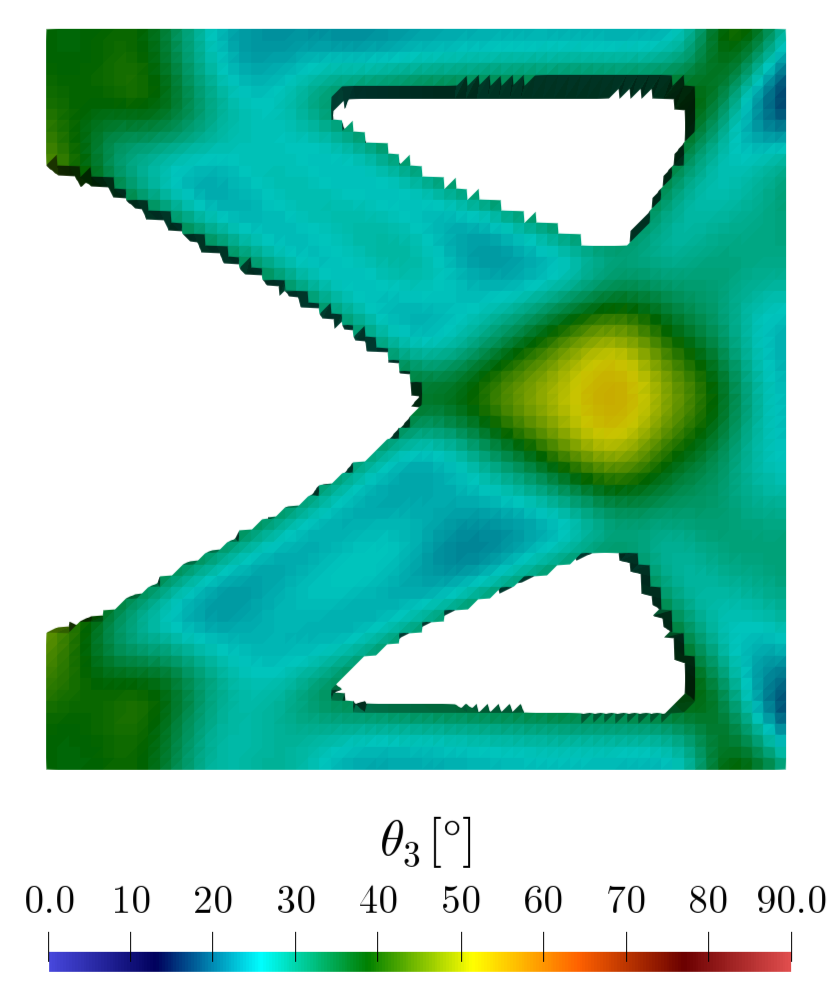}
		\caption{Optimized spinodoid parameter $\theta_3$}
	\end{subfigure}
	\qquad \qquad
	\begin{subfigure}{0.32\textwidth}
		\centering
		\includegraphics[width = \linewidth]{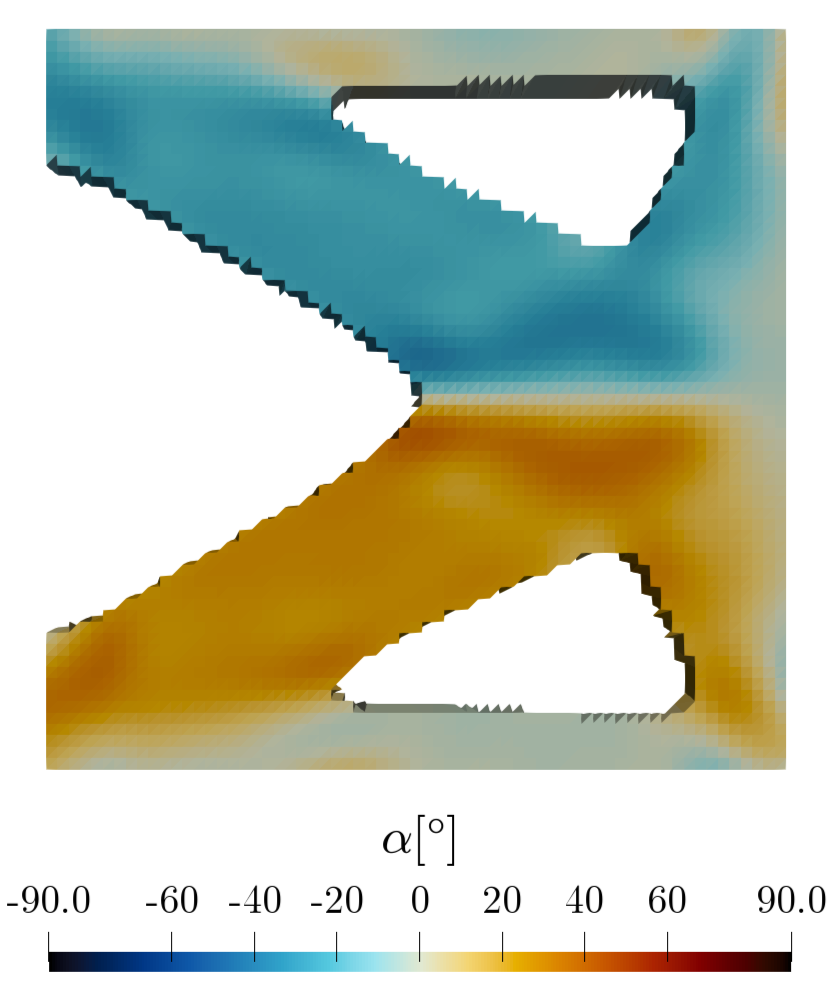}
		\caption{Optimized material orientation $\alpha$}
		\label{fig:multiSymmAlpha}
	\end{subfigure}
	\caption{\textit{Benchmark III:} Front view of the optimal topology (material distribution, anisotropy, and orientation) of the spinodoid-based structure, optimized for two symmetric load cases. Examples of the spatially-varying design parameters and the corresponding microscale topologies are shown in (a).}
	\label{fig:multiSymmResults}
\end{figure}

The optimized design obtained for spinodoid microstructures and SIMP are shown in Figures~\ref{fig:multiSymmResults} and \ref{fig:multiSymmSIMP}, respectively. Similar to previous benchmarks, the former achieves an improvement of \LZ{$4.88\%$} in compliance (\figurename\ref{tab:multiSymmObj}) over the SIMP-based design due to the spatially-varying material distribution, anisotropy, and orientation on the microscale.

\subsection{Benchmark IV: Multiple load cases -- non-symmetric loading}

\begin{figure}
	\centering
	\begin{subfigure}{0.5\textwidth}
		\centering
		\includegraphics[width = \textwidth]{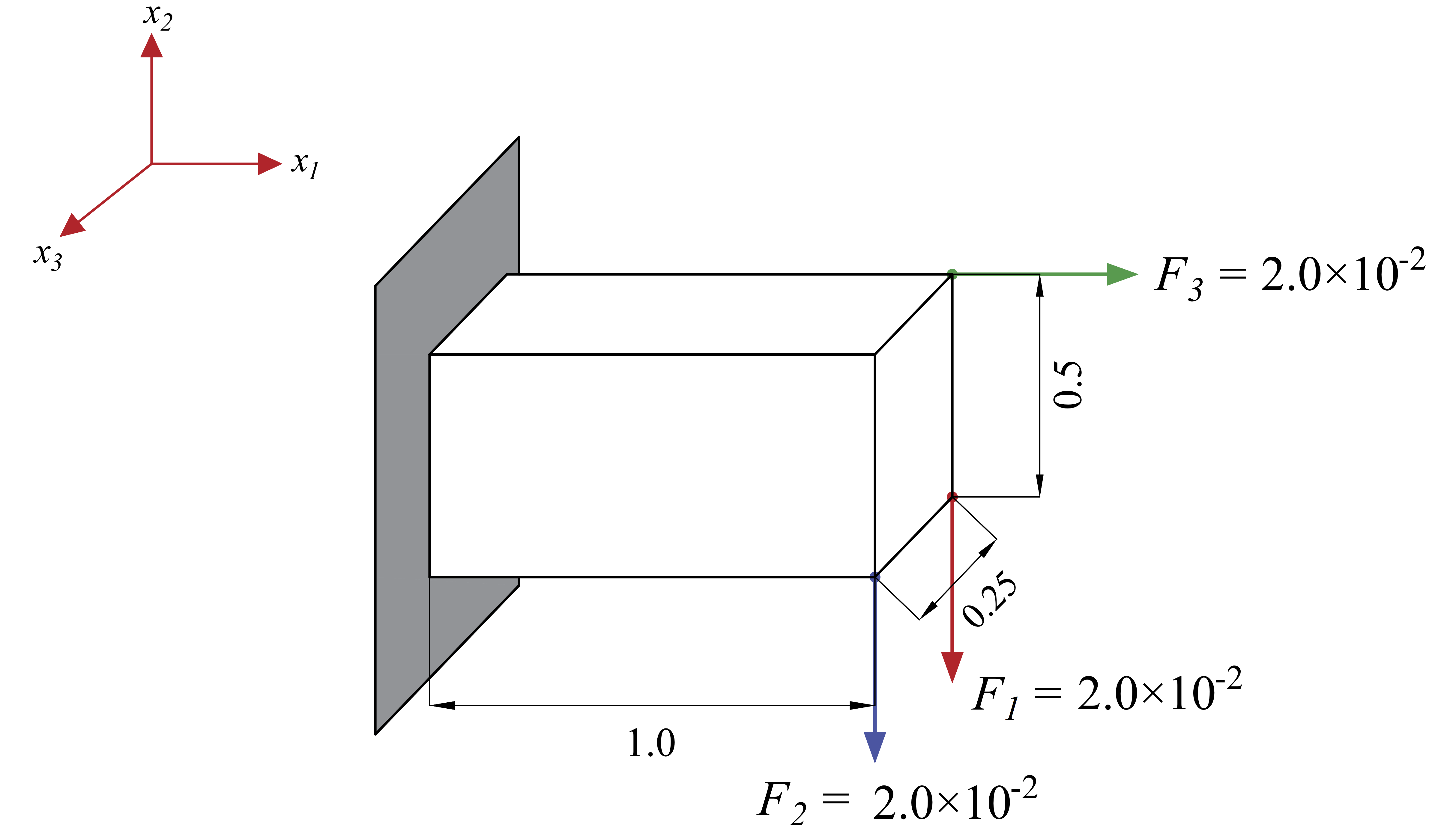}
		\caption{}\label{fig:multiNonSymm}
	\end{subfigure}
	\qquad
	\begin{subfigure}{0.45\textwidth}
		\centering
		\includegraphics[width =\textwidth]{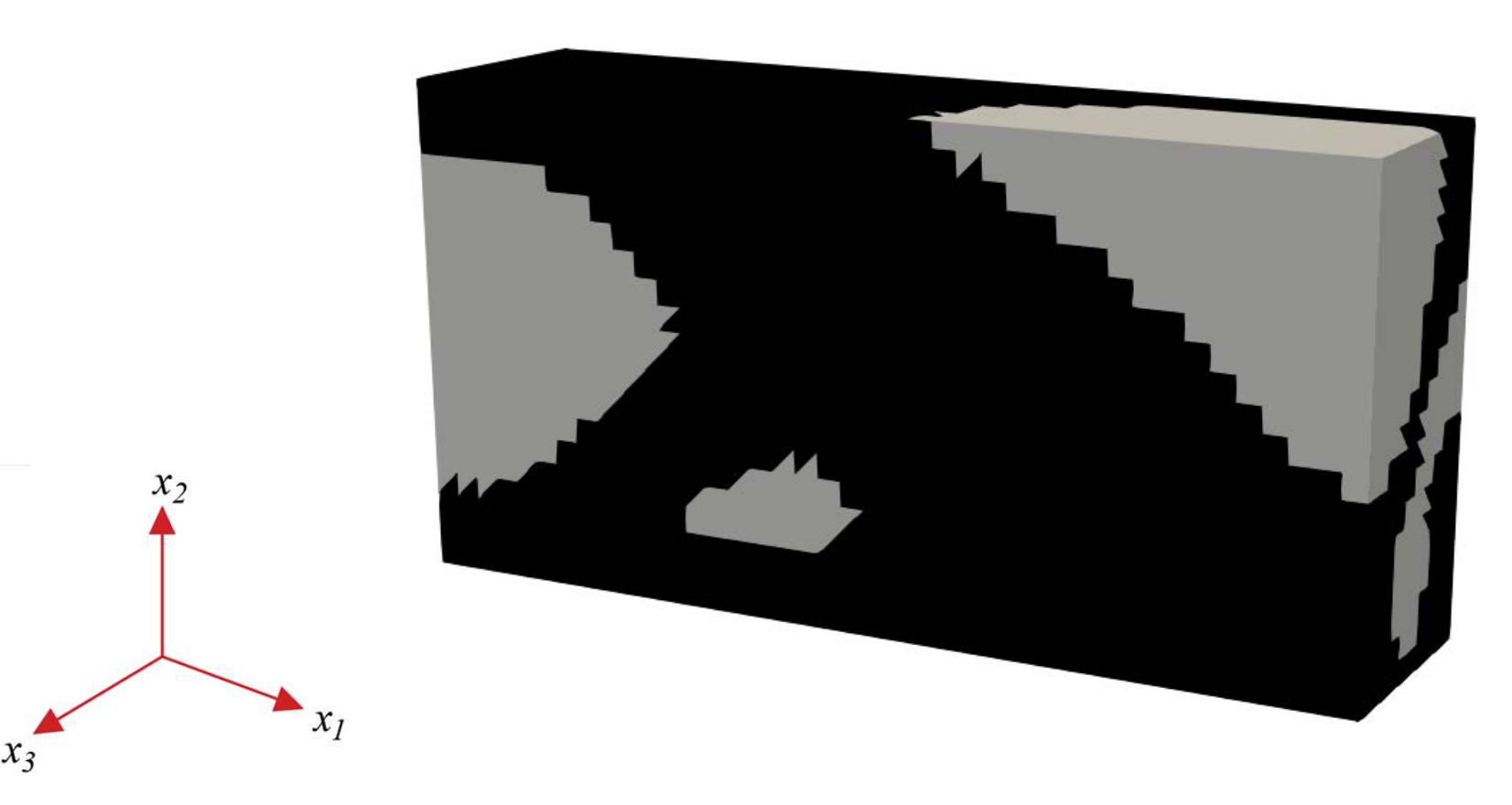}
		\caption{}
		\label{fig:multiNonSymmSIMP}
	\end{subfigure}

	\begin{subfigure}{0.59\textwidth}
		\centering
		\begin{tabular}{cccccccc}
			\toprule
			& \textbf{Spinodoid} & \textbf{SIMP} \\
			\midrule
			Optimized compliance & $7.8 \times 10^3$ & $7.9 \times 10^3$ \\
			\toprule
		\end{tabular}
		\caption{}
		\label{tab:multiNonSymmObj}
	\end{subfigure}

	\caption {\textit{Benchmark IV:} (a) Schematic of the structure to be optimized for three non-symmetric load cases. \LZ{(b) 3D view of the optimal topology obtained using the SIMP method. Gray regions here denote complete void.} (c) Comparison of the optimal compliance with spinodoid microscale architecture (via the proposed method) and solid material (via SIMP).}
\end{figure}

We modify \textit{Benchmark III} (Section~\ref{sec:Benachmark3}) by considering the three non-symmetric load cases shown in \figurename \ref{fig:multiNonSymm}, which involve three point loads of the same magnitude but at three different corners ($M=3$). An average volume constraint of $\bar\rho=0.4$ is imposed. The domain is discretized into a uniform linear tetrahedral mesh with 48,000 elements and 240,000 design variables.

\begin{figure}
	\centering
	\includegraphics[width = \linewidth]{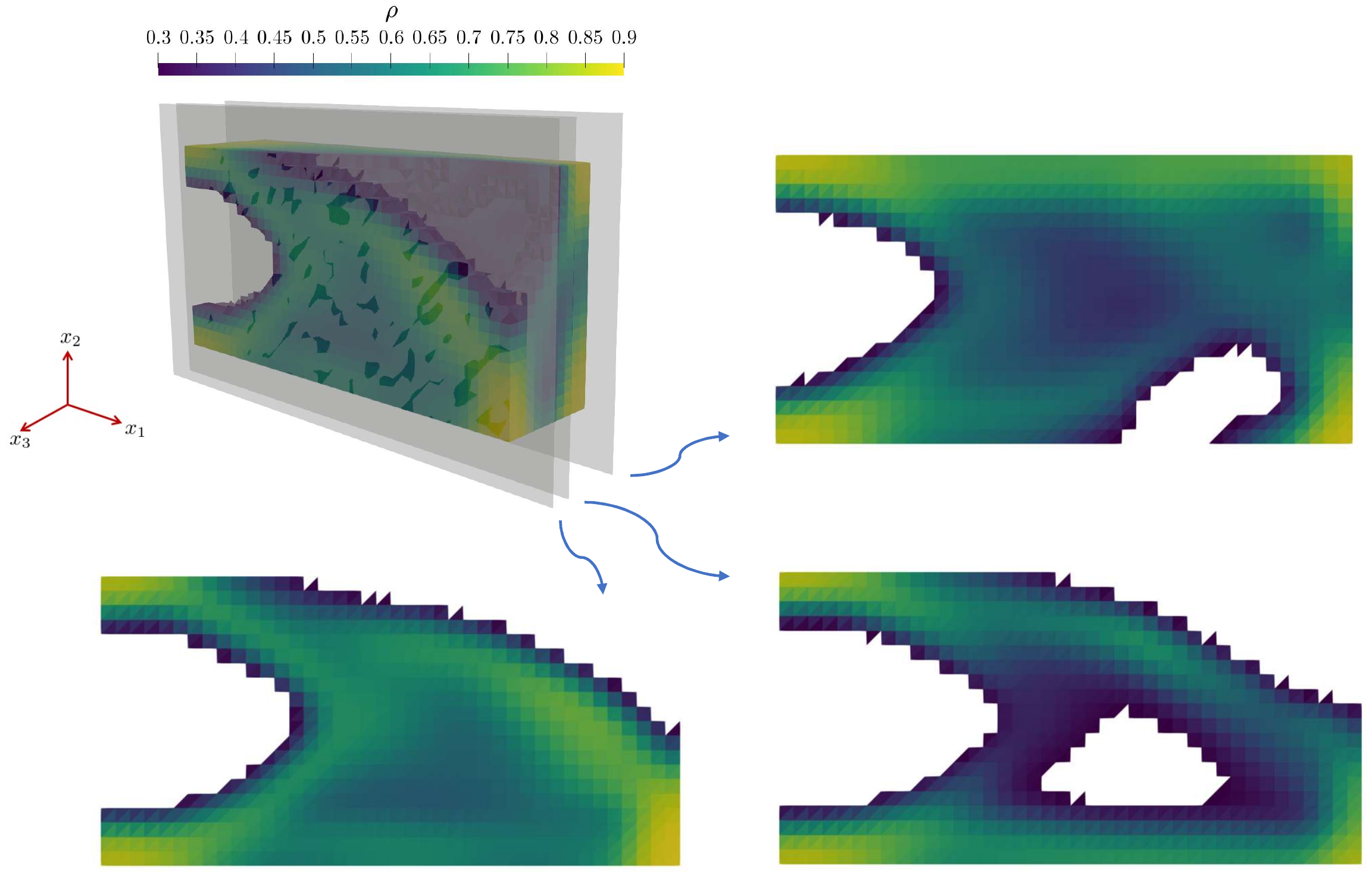}
	\caption{\textit{Benchmark IV:} Optimal material distribution in the spinodoid-based structure with multiple non-symmetric load cases. Front-view of multiple cross-sectional planes are shown here.}
	\label{fig:multiNonSymmrho}
\end{figure}

\begin{figure}
	\centering
	\begin{subfigure}{\textwidth}
		\centering
		\includegraphics[width =0.95 \linewidth]{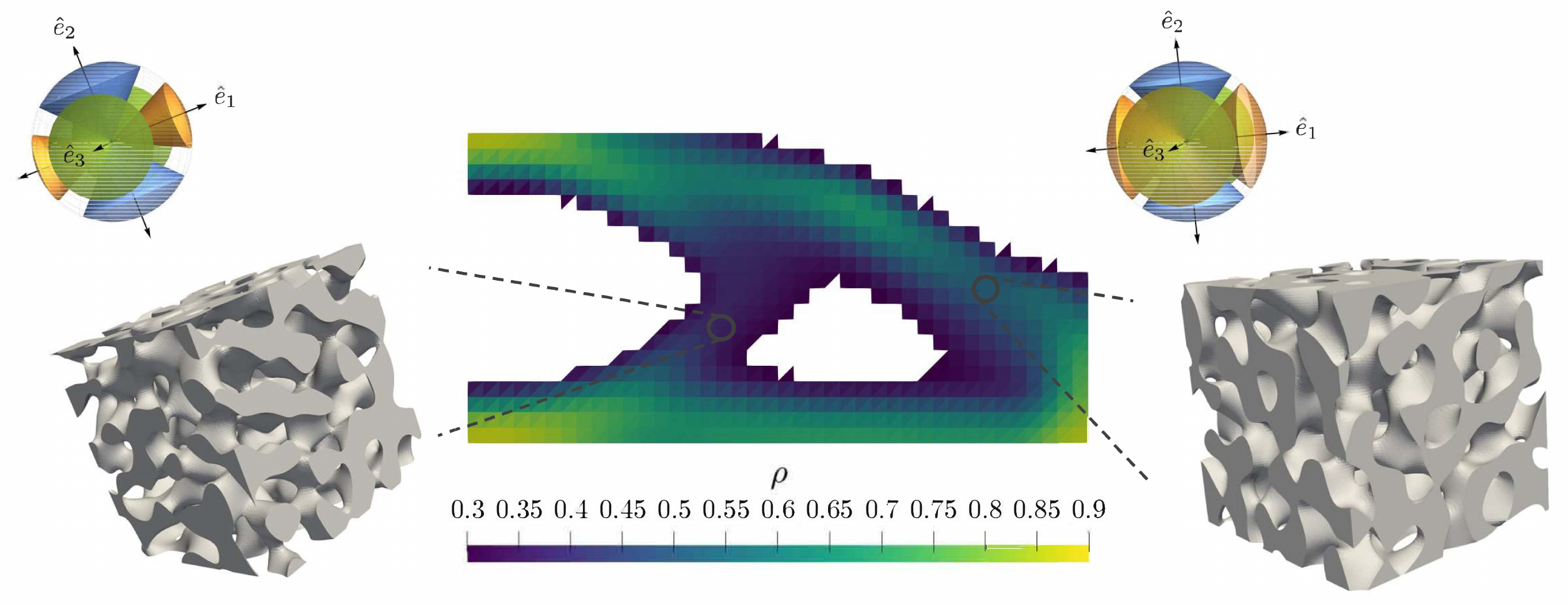}
		\caption{Optimized relative density: $\rho$}
		\label{fig:multiNonSymmRho}
	\end{subfigure}
	\begin{subfigure}{\textwidth}
		\centering
		\includegraphics[width = 0.95\linewidth]{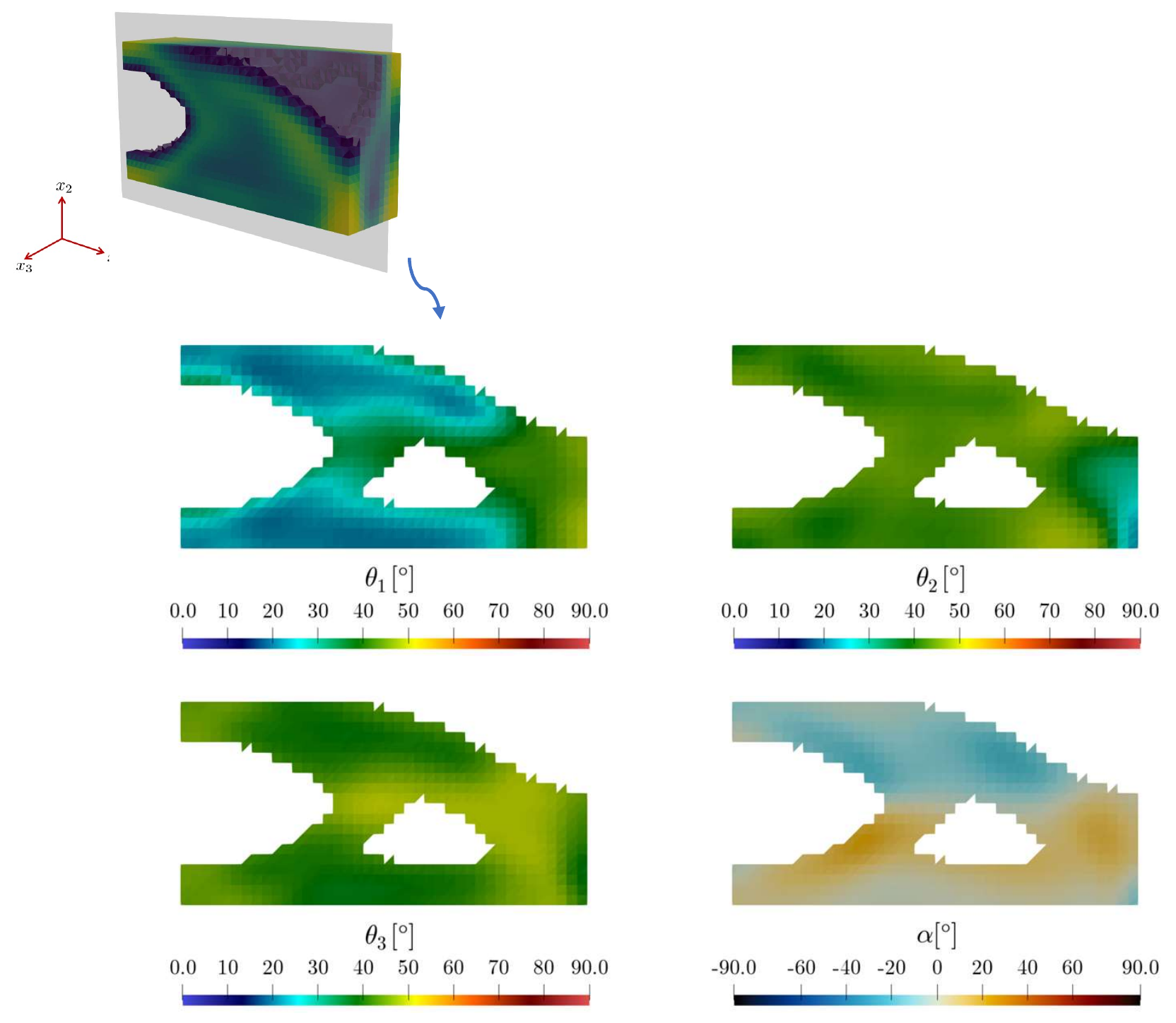}
		\caption{Optimized spinodoid parameter: $\theta_1, \theta_2, \theta_3, \alpha$}
	\end{subfigure}
	\caption{\textit{Benchmark IV:} Optimal topology (material distribution, anisotropy, and orientation) of the spinodoid-based structure optimized for three non-symmetric load cases. The front view of the cross-section at $x_3 = 0.2$ is shown here. Examples of spatially-varying design parameters and the corresponding microscale topologies are shown in (a).}
	\label{fig:multiNonSymmResults}
\end{figure}

Optimal SIMP- as well as spinodoid-based designs are illustrated in Figures~\ref{fig:multiNonSymmSIMP} as well as in Figures~\ref{fig:multiNonSymmrho} and \ref{fig:multiNonSymmResults}, respectively. Compared with the previous benchmarks, the spinodoid-based design shows only minor improvement (1.27\%) in compliance over SIMP (\figurename\ref{tab:multiNonSymmObj}). This can, in fact, be expected for the simultaneous optimization for multiple load cases applied in different directions, since -- under such constraints -- the structure is likely to favor isotropic topologies as the best compromise between all design cases, thus leading to similar performance as SIMP. We note that the small improvement of 1.27\% in compliance further needs to be considered with caution, as it can partially be due to numerical artifacts.



\section{Conclusions}
\label{sec:Conclusions}

We have presented a two-scale topology optimization framework for macroscopic bodies made of a spatially-variant microscale architecture based on spinodoid topologies. Inspired by microstructures produced by spinodal decomposition, the spinodoid topologies are anisotropic and described by a set of four design parameters (the relative density and the orientation distribution of wave vectors in the underlying Gaussian random field). The topology optimization problem minimizes the linear elastic compliance of macroscopic bodies by solving both for the displacement field and for a continuous field of the design parameters. The effective material response at any point on the macroscale is identified with the homogenized, effective response of a representative volume element filled by the spinodoid topology defined by the local design parameters. To bypass costly computational homogenization simulations, we here introduce a new approach based on a deep neural network as a surrogate model that maps the design parameters onto the effective fourth-order stiffness tensor. Aside from significantly speeding up calculations, this approach also provides exact sensitivities (required for gradient-based optimization) at low numerical costs. \revised{Although we here focused on \textit{solid} spinodoid topologies with relative densities larger than 30\%, one could alternatively use \textit{shell}-type spinodal architectures to reach significantly lower relative densities \citep{hsieh2019mechanical,portela2020extreme,kumar2020inverse}}. We presented four benchmarks of linear elastic compliance optimization for different macroscopic bodies experiencing single and multiple load cases, which demonstrate the applicability of the framework and highlight advantages over, e.g., the classical SIMP approach due to the enlarged design space available by optimizing both macro- and microscales. Although our study was specific to spinodoid architectures and simple macroscale boundary value problems, the presented approach is sufficiently general to extend to other microscale architectures and more complex boundary value problems.

\bibliographystyle{elsarticle-harv}
\bibliography{Bib}

\appendix

\section{Generating a spatially-variant spinodoid topology with fully resolved microstructure}\label{sec:spatiallyVariantAppendix}

The nonlinear optimization problem \eqref{eq:objective} yields the set of spatially-variant design parameters \mbox{$\calX=\{\bfchi^e,e=1,\dots,n\}$} at the quadrature points \mbox{$\{\bfx^e,e=1,\dots,n\}$} of the macroscale finite element mesh. To generate the fully resolved spinodoid microstructure of the optimized macroscale body, we must turn that quadrature-point information into a seamless spinodoid architecture across the macroscale body $\Omega$. Here, we consider a weighted superposition of multiple GRFs, each described by its design parameters in $\calX$. Let $q(\bfx,\bfx^e)$ be a Gaussian weight function defined by
\be
q(\bfx,\bfx^e) = \frac{\exp (-\kappa \| \bfx-\bfx^e \|^2)}{ Z(\bfx)}
\ee
with a length scale parameter $\kappa>0$ and
\be
Z(\bfx) = \sum_{e=1}^n \exp (-\kappa \| \bfx-\bfx^e \|^2)
\ee
ensuring partition of unity, i.e., $\sum_{e=1}^n q(\bfx,\bfx^e) = 1$. Let $\varphi^e(\bfx)$ denote the GRF described by the design parameters {$\bfchi^e = (\rho^e,\theta_1^e,\theta_2^e,\theta_3^e,\alpha^e)^\text{T}$} and the corresponding level set
\be
\varphi_0^e = \sqrt{2}\text{erf}^{-1}(2\rho^e - 1).
\ee
When considering a total of $n$ quadrature points, we define an interpolated GRF
\be\label{eq:grf_interpolation}
\bar \varphi(\bfx) = \sum_{e=1}^n q(\bfx,\bfx^e) (\varphi^e(\bfx) - \varphi^e_0)
\ee
with the global level set (for all points $\bfx \in \Omega$)
\be
\bar \xi(\boldface{x}) = \begin{cases}
1 \text{ (solid)}&\quad  \text{if }\quad \bar \varphi(\boldface{x}) \leq 0, \\
0 \text{ (void)}& \quad \text{if }\quad \bar \varphi(\boldface{x}) > 0.
\end{cases}
\ee
For a given point $\bfx \in \Omega$ in the proximity of $\bfx^e$, the exponential decay in $q(\bfx,\cdot)$ (for sufficiently large $\kappa$) ensures that $q(\bfx,\bfx^e)\approx 1$ and $\bar\varphi(\bfx) \approx \varphi^e(\bfx)$ with level set $\varphi^e_0$; i.e., the interpolated GRF approximates the individual GRF described by $\bfchi^e$. Elsewhere, the above yields a superposition of those GRFs of nearby quadrature points to generate a seamlessly varying spinodoid topology throughout the entire domain $\Omega$. The computational efficiency of the interpolation can be improved by reducing the summation in \eqref{eq:grf_interpolation} to only those terms with  $q(\bfx,\cdot)$ greater than a minimum cut-off value. We point out that the GRF wave number $\beta$ must be sufficiently large to ensure an effective separation of scales between the micro- and macroscales.

In order to spatially resolve the microstructure over $\Omega$, $\bfx$ can conveniently be sampled from a separate mesh with significantly higher resolution than the mesh used for FEM (thus keeping the FEM costs low but providing high-resolution architectures for visualization and part production). Note that we here do not consider fabrication constraints nor specifics of additive manufacturing\revised{, which is a topic of discussion, e.g., in~\cite{Zegard2016}}.

\section{\revised{Computational performance }}
\label{sec:performanceNNAppendix}

\revised{
The computational times and resourses for different tasks are listed in Table~\ref{table:computing}. Runtimes reported roughly approximate the duration for the model to run on a single CPU, which are intented to provide only a qualitative measure. The DNN model takes 0.001 seconds for each prediction, significantly less than FEM homogenization which takes around 5 minutes. The computational time for stiffness computations using DNN is considerably reduced by several orders of magnitude.

}
\begin{table}[ht]
\centering
\begin{tabular}[t]{cccccc}
\toprule
\textbf{Task} & \textbf{Software} & \textbf{Parallelization \& Hardware}& \textbf{Runtime}\\
\midrule
Stiffness computation using FEM &  In-house C++ FEM code  &  16 MPI cores$^\mathsection$ &5 minutes   \\
Stiffness computation via the DNN$^\dagger$ &  PyTorch in Python   &  CPU, no parallelization$^\star$ &      0.001 seconds   \\
Training the DNN &  PyTorch in Python   &  CPU, no parallelization$^\star$ &     10 minutes    \\
Benchmark I&   Python   &  CPU, no parallelization$^\mathsection$ &     3 hours    \\
Benchmark II&   Python   &  CPU, no parallelization$^\mathsection$ &     3.5 hours    \\
Benchmark III &  Python   &  CPU, no parallelization$^\mathsection$ &     4.5 hours    \\
Benchmark IV &  Python   &  CPU, no parallelization$^\mathsection$ &     8 hours    \\
\toprule
\end{tabular}
	\caption{\revised{Computational resources and runtimes for all tasks.  $^\dagger$Runtimes for the DNN are measured for one prediction on a single data sample. $^\star$Computations were carried out on four 12-core 2.2 GHz Intel Xeon E5-2650 processors and 256 GB of DDR3 memory at 2500 MHz. $^\mathsection$Computations were carried out on the Euler IV cluster of ETH Z\"{u}rich with two 18-core 2.7 GHz Intel Xeon Gold 6150 processors and 192 GB of DDR4 memory at 2666 MHz. Data on stiffness computation and DNN training have been adapted from \cite{kumar2020inverse}, who introduced spinodoid architectures without the use of topology optimization.}}
	\label{table:computing}
\end{table}

\end{document}